\documentclass[3p,onecolumn,10pt]{elsarticle}

\usepackage{amssymb}
\usepackage{amsmath}
\usepackage[version=4]{mhchem}
\usepackage{doi}
\usepackage{comment}
\usepackage[T1]{fontenc}
\usepackage[official]{eurosym}
\usepackage{csvsimple}
\usepackage[dvipsnames,svgnames,x11names,table]{xcolor}
\usepackage{soul}
\usepackage{float}			
\usepackage{graphicx}
\usepackage{booktabs}
\usepackage{tabularx}
 \usepackage[flushleft]{threeparttable}
 \usepackage{placeins}
\usepackage{ltablex}
\usepackage{multicol}
\usepackage{multirow}
\usepackage{etoolbox}
\usepackage{caption}
\usepackage{subcaption}
\usepackage{libertine}
\usepackage{siunitx}
\DeclareSIUnit\ton{t}
\DeclareSIUnit\oil{o}
\DeclareSIUnit\equivalent{e}
\usepackage{cleveref}
\crefname{chart}{Chart}{Charts}
\crefname{section}{Section}{Sections}
\crefname{figure}{Fig.}{Figs.}
\crefname{graph}{Graph}{Graphs}
\crefname{scheme}{Scheme}{Schemes}
\crefname{equation}{Eq.}{Eqs.}
\crefname{table}{Table}{Tables}
\crefname{chapter}{Chapter}{Chapters}
\crefname{appendix}{}{}
\usepackage{xurl}
\usepackage{hyperref}

\usepackage{rotating} 

\usepackage{lineno}

\journal{Sustainable Energy \& Fuels}

\begin{document}

\begin{frontmatter}



\title{Minimum-regret hydrogen supply chain strategies to foster the energy transition of European hard-to-abate industries}


\author[a]{Alissa Ganter}
\author[a]{Paolo Gabrielli}
\author[a,b]{Hanne Goericke} 
\author[a]{Giovanni Sansavini}

\affiliation[a]{organization={Institute of Energy and Process Engineering},
            city={Zürich},
            postcode={8092},
            country={Switzerland}}
            
\affiliation[b]{organization={RWTH Aachen University},
            city={Aachen},
            postcode={52062}, 
            state={North Rhine-Westphalia},
            country={Germany}}       

\begin{abstract}

Low-carbon hydrogen (\ce{H2}) is envisioned to play a central role in decarbonizing European hard-to-abate industries, such as refineries, ammonia, methanol, steel, and cement. To enable its widespread use and support the transition, low-carbon \ce{H2} supply chain (HSC) infrastructure is required. Mature and economically viable low-carbon \ce{H2} production pathways include steam methane reforming of natural gas coupled with carbon dioxide (\ce{CO2}) capture and storage, water-electrolysis from renewable electricity, biomethane reforming, and biomass gasification. However, uncertainties surrounding demand and feedstock availabilities hamper their proliferation. Therefore, this work investigates the impact of uncertainty in future \ce{H2} demands and biomass availability on the HSC design. The HSC is modeled as a network of \ce{H2} production and consumption sites that are interconnected by \ce{H2} and biomass transport technologies. \ce{CO2} capture, transport, and storage infrastructure is modeled alongside the HSCs. We determine the cost-optimal low-carbon HSC design by solving a linear optimization model, considering a regional spatial resolution and a multi-year time horizon from 2022 to 2050. We adopt a scenario-based uncertainty quantification approach and define discrete scenarios with varying \ce{H2} demands and biomass availability. Applying a minimum-regret strategy, we show that planning for sufficiently large low-carbon \ce{H2} production capacities (about 9.6\,Mt/a by 2030) is essential to flexibly scale up HSCs and accommodate larger \ce{H2} demands of up to 35\,Mt/a by 2050. Although biomass-based hydrogen production technologies are identified as the most cost-efficient low-carbon hydrogen production technologies, investments are not recommended unless the availability of biomass feedstocks is guaranteed. In this context, investments in SMR-CCS and electrolyzers often offer greater flexibility. Furthermore, we highlight the importance of \ce{CO2} capture, transport, and storage infrastructure in the transition, which is required across scenarios. In particular the availability of \ce{CO2} removal technologies determine the ability to realize the 2050 net-zero emissions target. 
\end{abstract}



\end{frontmatter}

\clearpage
\section{Introduction}
\label{sec:introduction}

The envisioned role of hydrogen (\ce{H2}) in the future energy system has changed significantly throughout the years \cite{IEA2019,VanDerSpek2022}. Nevertheless, interest in \ce{H2} remains high, and its potential to decarbonize industrial sectors is widely acknowledged \cite{AgoraAFRY2021,EuropeanCommissionRepower2022}. 
The European industrial sector is currently responsible for 752\,Mt (21\,\%) of the annual anthropogenic greenhouse gas (GHG) emissions \cite{EEA2023}. Major contributors are the cement industry (15\,\%), the iron and steel industry (14\,\%), and the chemical industry, which includes refineries, methanol, and ammonia production (18\,\%) \cite{EEA2022}. These industries are difficult to decarbonize as they inherently rely on carbonaceous feedstocks and high-temperature heat, and therefore, are often referred to as "hard-to-abate" industries. 

Efficiency improvements can reduce process emissions to an extent \cite{Paltsev2021,InsightsChemIndustry2022}, however, additional measures are required to achieve the impending emissions targets, such as the EU Fit for 55 target, which requires a 55\,\% emission reduction with respect to 1990, and the net-zero \ce{CO2} emissions target for 2050 \cite{EuropeanCommissionFit552021}. Hence, a shift to low-carbon feedstocks and energy carriers is required. In this context, \ce{H2}, produced with low \ce{CO2} emissions (i.e., low-carbon \ce{H2}) is viewed as a promising solution. In Europe, \ce{H2} qualifies as "low-carbon" if process emissions are below 4.42\,kg\textsubscript{CO\textsubscript{2}}/kg\textsubscript{H\textsubscript{2}} \cite{CertifHy2022}. 

About 9\,\% (8.2\,Mt) of the global \ce{H2} demand is currently (2022) produced and consumed in Europe \cite{IEAGlobalH2Review2023}, 85\,\% of which is used as a feedstock to produce ammonia, methanol and other chemicals, or in refineries \cite{FonsecaJoana2023}. Aside from the current use of \ce{H2} as a feedstock, low-carbon \ce{H2} has the potential to replace coal as a reducing agent in the steel-making process \cite{Terlouw2019}, to generate high-temperature heat required in the cement-making process \cite{EuropeanHydrogenBackbone2021}, or, combined with captured \ce{CO2}, to replace carbonaceous feedstocks in chemicals production (e.g., methanol and plastics) \cite{Gabrielli2023}.

However, the future demand for low-carbon \ce{H2} is deeply uncertain. Governmental institutions (e.g., \cite{EuropeanCommissionHydrogenStrategy2020, EuropeanCommissionCleanPlanet2018, Cihlar2020}) and energy consultancies (e.g., \cite{AgoraAFRY2021,Terlouw2019,Bruegel2021}) investigate the potential of low-carbon \ce{H2} for decarbonizing European hard-to-abate industries and provide demand estimates for 2030 and 2050. Their 2050 demand estimates range from 2.4\,Mt to 40\,Mt. To enable the widespread use of low-carbon \ce{H2}, a European supply chain (HSC) infrastructure is required \cite{Neumann2023,Kotek2023}. However, the infrastructure development is strongly dependenant on the projections and assumptions surrounding future low-carbon \ce{H2} demands \cite{OchoaRobles2018,Robles2020}. 

In general, low-carbon \ce{H2} can be produced starting from renewable electricity via water-electrolysis and from biomass via biomass gasification or biomethane reforming \cite{Gabrielli2020}. Furthermore, \ce{H2} production from natural gas via steam methane reforming coupled with \ce{CO2} capture and storage (CCS) can be considered low-carbon if \ce{CO2} capture rates exceed 90\,\% and leakage rates of natural gas supply chains below 0.2\,\% \cite{Bauer2022}. A comparison of the available low-carbon \ce{H2} production routes identifies biomass-based \ce{H2} production as the most cost-efficient alternative while offering large reductions in \ce{CO2} emissions; especially when coupled with CCS, it enables net-negative emissions \cite{Parkinson2019}. However, difficulties in biomass collection, a lack of infrastructure, and competing interests with other sectors may substantially reduce the amount of biomass that can be dedicated for low-carbon \ce{H2} production \cite{Nevzorova2019, Schnorf2021, Daiogolou2015}. Therefore, the uncertainty in biomass availability must be considered when planning low-carbon HSCs.

Energy system optimization models have proven to provide useful insights for energy planners and policymakers \cite{Pfenninger2014}. They are widely used to investigate HSCs, offering an integrated representation of \ce{H2} production and transport, and enabling the analysis of trade-offs between technology alternatives over long-term, multi-period time horizons \cite{Blanco2022}. However, a key challenge lies in the uncertainty associated with the input data of energy system optimization models \cite{Yue2018}, and accuracy assessments reveal that energy system optimization models systematically underestimate uncertainties \cite{Wen2023, Koomey2003, Craig2003}. Therefore, the uncertainty should be accounted for in the decision-making process to increase robustness and derive long-term policy recommendations \cite{Linstone2004, Yue2018}. 

\cite{Riera2023} perform an extensive literature review of existing HSC models. While most HSC models are deterministic, several works exist that include uncertainty, with \cite{Hugo2005,Almansoori2012} leading the way. Thus far, many studies pertain to the uncertainty in the \ce{H2} demand, e.g. \cite{Bique2022,Robles2020,Yang2020,Fazli-Khalaf2020}, which is identified as the most influential parameter for the HSC design \cite{OchoaRobles2018}.

Four approaches are commonly used to address the uncertainty in the input data: (1) Monte Carlo analysis (e.g., \cite{Yang2020}), (2) stochastic programming (e.g., \cite{Almansoori2012,Nunes2015}), (3) robust optimization (e.g., \cite{Lou2014}), and (4) scenario-based uncertainty analysis (e.g., \cite{De-LeonAlmaraz2014}). While (1)-(3) provide clear recommendations to decision-makers by directly accounting for the uncertainty in the input data, they also largely increase the model complexity making it difficult to maintain feasibility. Considering computational limitations, scenario-based approaches are often better suited to include uncertainty in large-scale energy system models \cite{Yue2018}. Min-max regret criteria can be used to hedge against parameter variations and identify the solution that performs best, even in the worst case \cite{Aissi2009}. 

To the best of our knowledge, no uncertainty analysis of the European HSC infrastructure rollout exists to date. This work aims to address this gap and identifies minimum-regret strategies that result in the lowest costs considering the deep uncertainty surrounding the future \ce{H2} demand and the availability of biomass feedstocks for \ce{H2} production. In particular, we investigate (1) how uncertainties in the future \ce{H2} demand and biomass feedstock availability influence the optimal design and rollout of HSC and \ce{CO2} infrastructures, and (2) what \ce{H2} production technologies, feedstocks, and energy sources are consistently deployed in the optimum HSC infrastructure design of the future.

The paper is structured as follows. \cref{sec:method} describes the considered system (\cref{subsec:system_description}), the uncertainty quantification approach for \ce{H2} demand and biomass availability (\cref{subsec:uncertainty_analysis_hydrogen} and \cref{subsec:uncertainty_analysis_biomass}, respectively), and the solution strategy used to investigate the optimal HSC under uncertainty and identify the minimum-regret strategy (\cref{subsec:optimization_model} and \cref{subsec:minimum_regret_solution}, respectively). \cref{sec:results} presents the results and \cref{sec:discussion} discusses their implications. Finally, \cref{sec:conclusion} draws conclusions.

\section{Optimal design of hydrogen supply chains (HSCs) under uncertainty}
\label{sec:method}

\begin{figure}[h!]
    \centering
    \includegraphics[width=\textwidth]{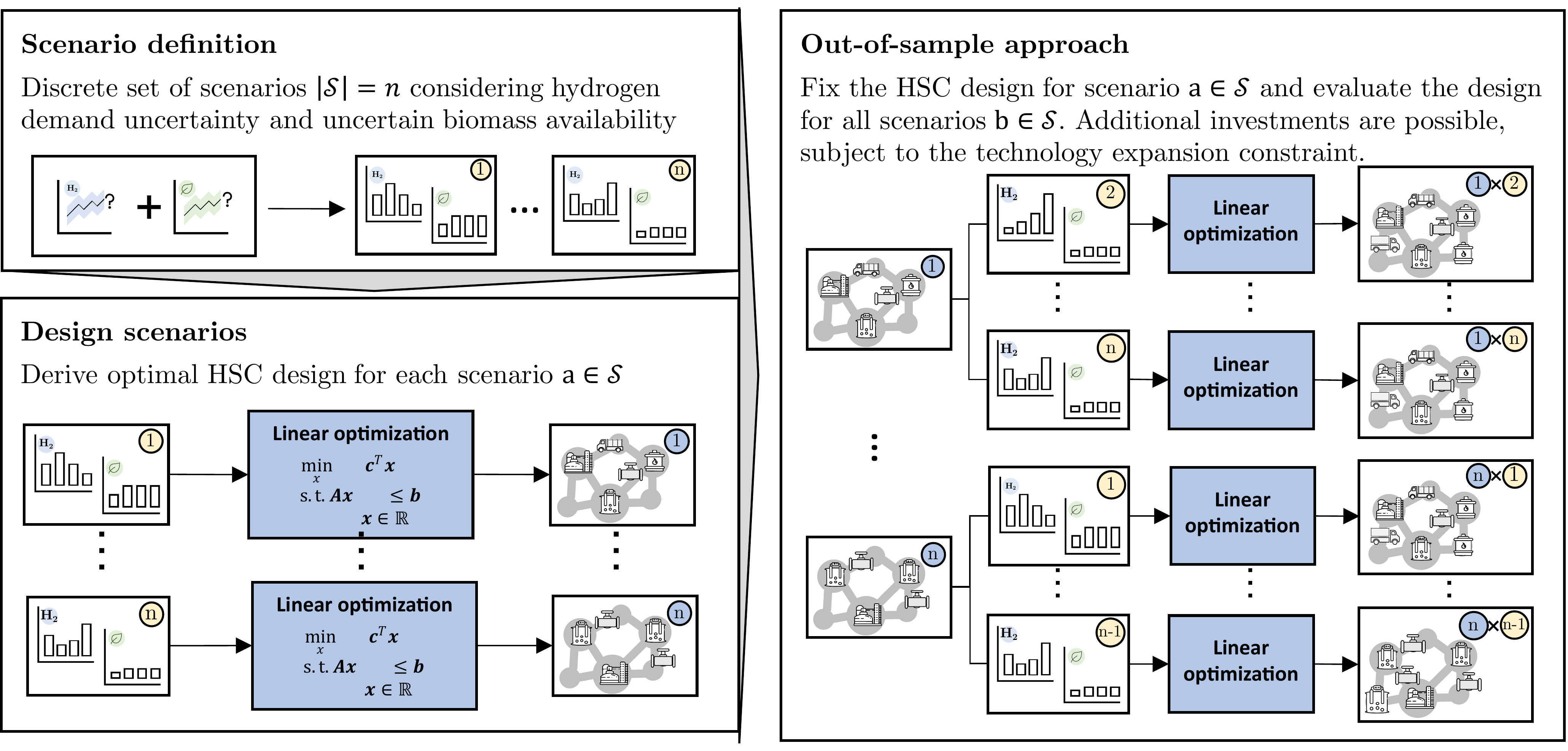}
    \caption{Solution strategy adopted to identify the minimum-regret \ce{H2} supply chain design considering uncertainties in the evolution of the future \ce{H2} demand and in the availability of biomass.}
    \label{fig:solution_strategy}
\end{figure}

The elements of the analyzed HSCs are described in \cref{subsec:system_description}. Furthermore, \cref{fig:solution_strategy} visualizes the methodology developed to identify the minimum-regret HSC design considering uncertainties in future \ce{H2} demands and biomass availability. "Scenario definition" derives a discrete set of scenarios $\mathcal{S}$ by combining the \ce{H2} demand and biomass availability projections (\cref{subsec:uncertainty_analysis_hydrogen,subsec:uncertainty_analysis_biomass}). "Design scenario" determines the optimal HSC design for each scenario $a \in \mathcal{S}$ using a deterministic, linear optimization model outlined in \cref{subsec:optimization_model}. The resulting HSC design describes the optimal low-carbon \ce{H2}, \ce{CO2}, and biomass infrastructure rollout for a given \ce{H2} demand and biomass availability; allowing governments to deduce policies promoting low-carbon technologies and incentivizing manufacturers to ramp up their capacities to enable the rapid scale-up of low-carbon technology capacities and infrastructure. "Out-of-sample approach" evaluates the HSC designs derived in "Design scenario", when operated under all other possible scenarios $b \in \mathcal{S}, b \neq a$. Additional investments may be needed to adapt the initial supply chain design derived under scenario $a$ to meet the target decarbonization pathway under the new conditions of scenario $b$. However, these additional investments, and thus, the speed at which technology capacities can be expanded, are limited by the existing and planned capacities of the technology manufacturers \cite{Leibowicz2016}. We include this by adding a technology expansion constraint to the optimization model, which estimates the speed of the technology expansion based on the existing capacity stocks (\cref{subsec:optimization_model} and \cref{supp:technology_expansion}). 
Finally, the performance of each design scenario of the HSC is evaluated based on the levelized cost of \ce{H2} (LCOH) of the out-of-samples scenarios. The minimum-regret solution is the design scenario, for which (i) the highest cost of the out-of-sample scenario is the lowest across all design scenarios (i.e. the min-max LCOH) and (ii) all out-of-sample scenarios meet the annual emission targets (\cref{subsec:minimum_regret_solution}).

\subsection{System description}
\label{subsec:system_description}

\begin{figure}[!htp]
    \centering
    {\noindent\includegraphics[width=0.85\textwidth]{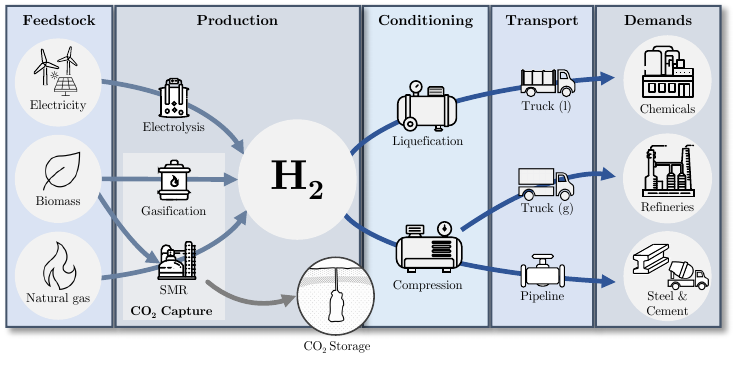}}
    \caption{Overview of the available feedstocks and energy sources, the available \ce{H2} production, conditioning, and transport technologies, and the considered \ce{H2} demands.}
    \label{fig:system_design}
\end{figure}

The HSC is modeled as a network of nodes and edges, with regional resolution following the EU's Nomenclature of territorial units for statistics for level 2 (NUTS 2) \cite{NUTS22021}. At each node, \ce{H2} can be produced through a portfolio of feedstocks and energy sources, including natural gas, electricity, and biomass. To reduce \ce{CO2} emissions, \ce{H2} production from natural gas and biomass can be coupled with \ce{CO2} capture technologies. Edges connect the nodes. The distance between two nodes is approximated by their Haversine distance. At each edge, \ce{H2}, \ce{CO2}, and biomass transport technologies can be installed. In the following, the individual components of the supply chain are described in more detail. \Cref{fig:system_design} provides an overview of the available feedstocks and energy sources, the available \ce{H2} production and transport technologies, and the considered \ce{H2} demands. The full model description is published in~\cite{Ganter2024}.

\paragraph{Feedstocks and energy sources.} The considered feedstocks and energy sources are natural gas, electricity, and biomass. We assume that natural gas and grid electricity are available at each node. Furthermore, renewable electricity can be generated from wind and solar energy. The wind and solar energy potentials are modeled as the technical potentials reported in \cite{Ruiz2019,Gonzalez-Aparicio2021}. Furthermore, the availability of biomass is subject to uncertainty and limited according to the selected scenario (\cref{subsec:uncertainty_analysis_biomass}). 

\paragraph{\ce{H2} production technologies} The available \ce{H2} production technologies are (i) steam methane reforming (SMR) from natural gas or biomethane, (ii) water-electrolysis from electricity, and (iii) biomass gasification. SMR and biomass gasification can be coupled with \ce{CO2} capture and storage to lower process emissions. The techno-economic parameters of the \ce{H2} production technologies are reported in \cref{supp:input_data_technologies}.

\paragraph{\ce{H2} transport technologies} The available \ce{H2} transport technologies are trucks and pipelines. The transport conditions for \ce{H2} vary depending on the transport mode. Trucks transport \ce{H2} in ISO-tank containers (isotainers) as a compressed gas or in its liquid form (\ce{H2} truck gas and \ce{H2} truck liquid), and pipelines transport \ce{H2} as a compressed gas. To meet the specific transport requirements, \ce{H2} is conditioned (i.e., compressed or liquefied). The techno-economic parameters of the \ce{H2} transport and conditioning technologies are reported in \cref{supp:input_data_technologies}.

\paragraph{\ce{H2} demand} Current (2020) and potential future \ce{H2} demands for ammonia production, methanol production, refineries, steel production, and cement production are considered. To account for the uncertainty in future \ce{H2} demands, different \ce{H2} demand scenarios are investigated (\cref{subsec:uncertainty_analysis_hydrogen} and \cref{supp:regional_hydrogen_demand_uncertainty}).

\paragraph{\ce{CO2} supply chain} A \ce{CO2} supply chain is designed alongside the \ce{H2} supply chain to account for the transport of the \ce{CO2} captured at the production sites to the storage locations. We consider 27 potential \ce{CO2} storage locations in Europe following \cite{IOGP2023}. The \ce{CO2} storage capacity is limited by the capacity of existing and announced \ce{CO2} storage projects, which corresponds to 132\,Mt/a \cite{IOGP2023}. The available \ce{CO2} transport technologies are trucks (with isotainers) and pipelines. Similarly to \ce{H2}, the captured \ce{CO2} is conditioned to meet transport requirements. The techno-economic parameters for \ce{CO2} transport are reported in \cref{supp:input_data_technologies}.

\paragraph{Biomass supply chain} Two types of biomass are considered: dry and wet biomass. Dry biomass consists of woody biomass, which can be transported for long distances via containers loaded on trucks \cite{Ruiz2015,Schnorf2021}. Wet biomass consists of manure and waste biomass, whose collection and transport is challenging \cite{Schnorf2021}, with maximum transport distances ranging between \SIrange{10}{50}{\kilo\meter} \cite{Schnorf2021,Scarlat2018}. Therefore, we assume that wet biomass cannot be transported and is only used at the locations where it is available. Wet biomass serves as feedstock for anaerobic digestion to produce biomethane, which can be transported via isotainers loaded on trucks \cite{ZorgBiogasGmbH} or injected into the natural gas grid \cite{Terlouw2019}. We assume that the gas grid is available at each node of the supply chain, and that biomethane can be injected into the gas grid subject to grid connection costs \cite{Terlouw2019}. The techno-economic parameters for dry biomass transport and biomethane transport are reported in \cref{supp:input_data_technologies}. 

\subsection{Hydrogen demand uncertainty}
\label{subsec:uncertainty_analysis_hydrogen}

The uncertainty analysis covers \ce{H2} demand predictions for European hard-to-abate industries, namely refineries, ammonia, methanol, steel, and cement industries. A literature search is conducted to collect \ce{H2} demand forecasts for the different industries and covers publications from 2018 to today. In total, 28 literature scenarios are analyzed (\cref{tab:hydrogen_demand_scenarios}). Scenarios that do not provide a breakdown of the \ce{H2} demand estimates for the considered industries are excluded from further analysis. 

The feasibility and attractiveness of hydrogen-based solutions is influenced by technological, economic, and political factors. Uncertainties surrounding technological factors such as technological breakthroughs and the time required for commercialization, as well as uncertainty surrounding economic factors such as the cost-competitiveness of low-carbon \ce{H2} technologies and the uncertainty surrounding future cost trajectories can hinder their adoption. In contrast, political measures such as \ce{CO2} pricing mechanisms can address affordability issues and promote investments \cite{Blanco2022,Schlund2022}. 

In particular, assumptions surrounding (1) material and process efficiency, (2) electrification, (3) recycling, and (4) \ce{CO2} capture and storage strongly influence the \ce{H2} demand predictions \cite{MaterialEconomics2019}. For example, in ammonia production, material and process efficiency improvements could reduce \ce{H2} feedstock requirements by up to 25\,\% with respect to today's values \cite{MaterialEconomics2019}. In steel industry, recycled (or secondary) steel could replace 50-70\,\% of the primary steel production, thereby reducing potential future \ce{H2} demands substantially \cite{Luh2020}. The availability of electricity-based decarbonization pathways and the option of \ce{CO2} capture and storage add further layers of uncertainty \cite{Azadnia2023}. 

\begin{figure}[!ht]
	\centering
{\noindent\includegraphics[width=0.8\textwidth]{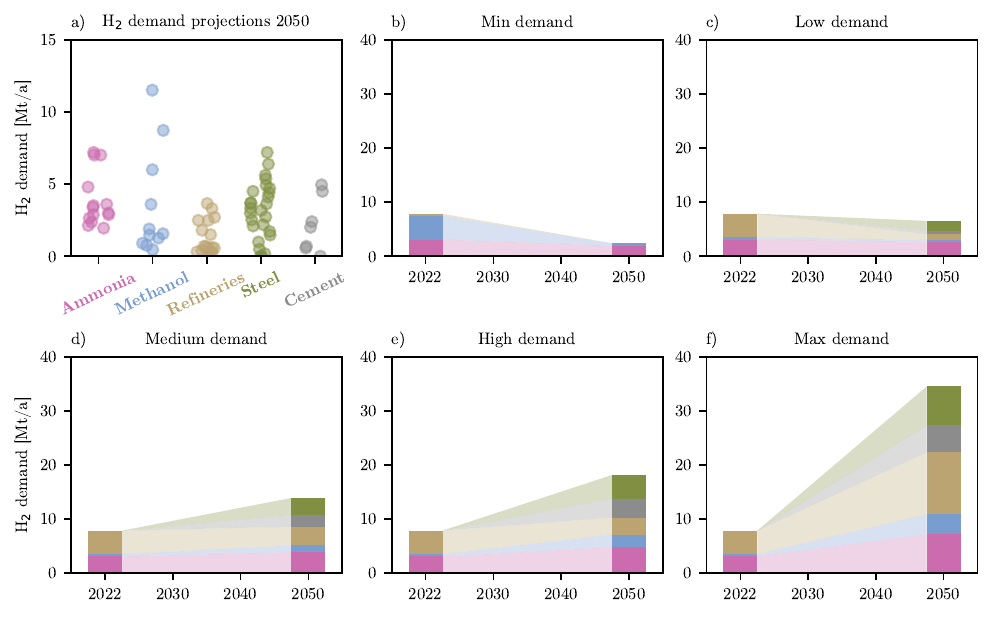}}
	\caption{a) European \ce{H2} demand estimates reported in the literature scenarios for refineries, ammonia, methanol, steel, and cement industry in 2050. b)-f) European \ce{H2} demand per industry in the minimum, low, medium, high, and maximum \ce{H2} demand scenario.}
	\label{fig:hydrogen_demand_scenarios}
\end{figure}

\cref{fig:hydrogen_demand_scenarios}a visualizes the 2050 \ce{H2} demand estimates for the considered industries reported across literature (\cref{tab:hydrogen_demand_scenarios}). In general, the spread of projected \ce{H2} demands is lower in industries that inherently rely on \ce{H2} as a feedstock in their production process (ammonia production plants and refineries). The spread increases for industries where multiple decarbonization strategies compete with each other (e.g., steel and cement industry). The largest spread is observed for methanol, an important base chemical that is required e.g., for plastics production via the methanol-to-olefins route. While \cite{Terlouw2019,EuropeanHydrogenBackbone2021} expect a large uptake of methanol-to-olefins, and thus, the methanol demand, \cite{AgoraAFRY2021} provide more conservative methanol demand estimates. \Cref{supp:regional_hydrogen_demand_uncertainty} provides a detailed analysis of the variability in the regional \ce{H2} demand estimates.

\begin{table}
\centering
\begin{tabularx}{\textwidth}{XccXc}
    \toprule
    \textbf{Title} & \textbf{No. Scenarios} & \textbf{2050 demand [Mt/a]} & \textbf{Industries} & \textbf{Source} \\ \midrule
    12 insights on hydrogen & 2 & 6.6-14.5 & Chemicals, Steel, Refineries & \cite{Fils2021}  \\ \midrule
    No-regret hydrogen & 1 & 8.1 & Chemicals, Steel, Refineries & \cite{AgoraAFRY2021} \\ \midrule
    Navigating through hydrogen & 3 & 8.0-18.6 & Chemicals, Steel, Refineries & \cite{Bruegel2021} \\ \midrule
    Clean planet for all & 11 & 0-18.5 & Chemicals, Steel, Refineries, Cement & \cite{EuropeanCommissionCleanPlanet2018} \\ \midrule
    Hydrogen roadmap  & 2 & 2.5-5.6 & Steel & \cite{Hebling2019}  \\ \midrule
    The optimal role for gas in a net zero emissions energy system & 1 & 9.9 & Chemicals, Steel, Refineries, Cement & \cite{Terlouw2019}  \\ \midrule
    Analysing future demand, supply, and transport of hydrogen & 1 & 733 & Chemicals, Steel, Refineries, Cement & \cite{EuropeanHydrogenBackbone2021}  \\ \midrule
    The potential of hydrogen for decarbonising EU industry & 2 & 6.3-18.8 & Chemicals, Refineries, Steel & \cite{Wachsmuth2021}  \\ \midrule 
    Hydrogen roadmap europe  & 2 & 12.2-19.6 & Chemicals, Steel, Refineries, Cement & \cite{HydrogenRoadmap2019}  \\ \midrule
    Industrial transformation 2050  & 3 & 5.6-10.7 & Chemicals, Steel, Cement & \cite{MaterialEconomics2019}  \\ \midrule
    \bottomrule
\end{tabularx}
\caption{Hydrogen demand projections in literature for hard-to-abate industries in 2050.}
\label{tab:hydrogen_demand_scenarios}
\end{table}

Five \ce{H2} demand scenarios are developed to tackle the deep uncertainty associated with the future \ce{H2} demands: minimum (min), low, medium, high, and maximum (max). The medium demand scenario represents the average of the \ce{H2} demand estimates. The low and high demand scenarios are derived based on the 25\textsuperscript{th} and 75\textsuperscript{th} quartile of the \ce{H2} demand estimates. Quartiles are used here because they are insensitive to outliers, but maintain the information about the center and spread of the \ce{H2} demand estimates \cite{Krzywinski2014}. 
Finally, the min and max demand scenarios represent the minimum and maximum \ce{H2} demand estimates, and are added to cover the full range of the \ce{H2} demand estimates. \cref{fig:hydrogen_demand_scenarios}b-f visualize the temporal evolution of the five \ce{H2} demand scenarios, where data is collected for 2020 and 2050, and linear interpolation is used for intermediate years.
 
\subsection{Biomass availability uncertainty}
\label{subsec:uncertainty_analysis_biomass}
\cite{Ruiz2015} and \cite{Pana2021} estimate the technical and sustainable potential, respectively, of bioenergy for different types of biomass in Europe. However, the availability of biomass feedstocks is highly uncertain, and sustainability and socio-political factors such as the competition of biomass feedstocks with food production or alternative land uses are often not accounted for \cite{Speirs2015}. Here, we focus on sustainable biomass potentials, i.e. biomass that is not primarily grown for energy use and does not compete with food production or alternative land uses. Sustainable biomass includes residues from agriculture and forests and animal manure \cite{Wu2023}. 

To account for the uncertainty in biomass availability, we define three discrete scenarios: a reference scenario, a scenario with reduced biomass availability to account for competition with other sectors, and a scenario where no biomass is available for low-carbon \ce{H2} production. In the reference scenario, the estimates from \cite{Pana2021} are used. In the scenario with reduced biomass availability, we assume that the sectoral biomass consumption is proportional to the sectoral primary energy demands. According to the EU reference scenario, 26\,\% of the total energy consumption can be attributed to industry. Thus, in the reduced biomass scenario, only 26\,\% of the sustainable biomass potential is available for low-carbon \ce{H2} production. 

\subsection{Optimization model}
\label{subsec:optimization_model}

The optimal design of the HSC is determined via a linear optimization problem following \cite{Ganter2024}. In general form, a linear optimization problem can be formulated as follows:
\begin{equation}
\begin{split}
\min_{\mathbf{x}} & \quad \mathbf{c}^\mathrm{T} \mathbf{x}  \\
\text{s.t.} & \quad \mathbf{A x} \leq \mathbf{b} \\
& \quad \mathbf{x}\in\mathbb{R}^N
\end{split}
\end{equation}
where the objective function is expressed as a linear combination of continuous decision variables $\mathbf{x}$ with dimension $N$ and coefficients $\mathbf{c}$; and the constraints are expressed as a linear combination of matrix $\mathbf{A}$, decision variables $\mathbf{x}$, and vector $\mathbf{b}$.

The optimization problem is implemented in the optimization framework ZEN-garden (Zero-emissions Energy Networks), developed at the Reliability and Risk Engineering Lab at ETH Zurich. ZEN-garden optimizes the design and operation of energy system models to investigate transition pathways toward decarbonization \cite{Ganter2024,Mannhardt2023b,Mannhardt2023}. The optimization problem is solved using the commercial solver Gurobi \cite{GurobiOptimizationLLC2022}.

\paragraph{Input Data}
The input data to the optimization problem includes (i) spatially-resolved \ce{H2} demands, carrier prices, \ce{CO2} intensities, and availabilities of biomass, wind, and solar energy, (ii) the techno-economic parameters describing the cost and performance of production, conditioning, and transport technologies, (iii) the existing \ce{H2} production capacities, (iv) the size and location of the available \ce{CO2} storage sites, and (v) the target decarbonization pathway. A yearly resolution is used to model the time-dependent variables, namely the \ce{H2} demands, carrier prices, \ce{CO2} intensity of the electricity grid, and biomass availability. The input data is reported in \cref{supp:input_data}.

\paragraph{Decision Variables}
The optimization problem determines (i) the optimal selection, capacity, and location of the \ce{H2} production, conditioning, and transport technologies, (ii) the energy inputs and outputs of each \ce{H2} production and conditioning technology, (iii) the carrier flows through each transport technology, (iv) the nodal carrier imports and exports.

\paragraph{Constraints}
The constraints of the optimization problem include (i) the nodal mass balances for electricity, natural gas, \ce{H2}, biomass, and \ce{CO2}, (ii) the performance and operating limits of the \ce{H2} production, conditioning, and transport technologies, and (iii) the \ce{CO2} emissions constraint limiting the yearly \ce{CO2} emissions to the target emissions values of the selected decarbonization pathway. Here, we assume linearly decreasing \ce{CO2} emissions from today's values such that the 2050 net-zero emissions (NZE) target is achieved \cite{EuropeanCommissionClimateLaw2021}. To ensure the feasibility of the optimization problem, a slack variable is added to the \ce{CO2} emission constraint. This slack variable can be interpreted as a \ce{CO2} emissions overshoot of the emission target and is associated with a large cost (100k€/
t), ensuring that overshooting the \ce{CO2} emission constraint is selected as a last resort. The HSC design is considered a feasible solution if the \ce{CO2} emissions target is achieved and the \ce{CO2} emissions overshoot is zero.

The constraints are detailed in \cite{Ganter2024}. Here, we expand the model formulation from \cite{Ganter2024} by adding technology expansion constraints when performing the out-of-sample approach (\cref{fig:solution_strategy}). The technology expansion constraints limit the maximum annual growth rate of a technology based on the existing capacity stock and are formulated following \cite{Leibowicz2016,Mannhardt2023} and described in \cref{supp:technology_expansion}. The technology expansion is parameterized based on historically observed growth rates. \cite{Mannhardt2023} investigate the historical annual growth rates for low-carbon technologies and observe growth rates between 10\,\% (wind offshore) to 29\,\% (solar PV). For our reference case, we select a technology expansion rate of 20\,\%. A sensitivity analysis is performed for technology expansion rates varying from 10\,\% to 29\,\%.

\paragraph{Objective function}
The optimization problem minimizes the net present cost of the system, which includes the investment and operating costs of the \ce{H2}, \ce{CO2}, and biomass supply chains in compliance with the \ce{CO2} emission targets.

\subsection{Minimum-regret solution}
\label{subsec:minimum_regret_solution}
The minimum regret solution is the HSC scenario design that (i) results in the lowest levelized cost of \ce{H2} (LCOH) for the worst-case out-of-sample scenario (i.e., the min-max across the LCOH of the out-of-sample scenarios), and (ii) is feasible for all scenarios. The LCOH is computed as the net present costs of the supply chains divided by the net present \ce{H2} production from 2022 to 2050; and feasibility is defined as the ability to meet the annual \ce{CO2} emissions targets, i.e. the \ce{CO2} emission overshoot is zero.

\section{Results}
\label{sec:results}
The combination of five \ce{H2} demand levels (\cref{subsec:uncertainty_analysis_hydrogen}) and three biomass availabilities (\cref{subsec:uncertainty_analysis_biomass}) produces 15 scenarios. \Cref{subsec:optimal_design_base_scenarios} analyzes the optimal investment strategies across scenarios. \Cref{subsec:optimal_design_base_scenarios} contrasts the \ce{H2} and \ce{CO2} transport infrastructure requirements, and \Cref{subsec:LCOH} compares the levelized cost of \ce{H2} (LCOH) across the 15 design scenarios and evaluates their ability to fulfill the decarbonization pathway if one of the 14 other scenarios materializes, using the out-of-sample approach (\cref{subsec:LCOH}). Finally, \cref{subsec:minimum_regret_solution} discusses the characteristics of the minimum-regret solution.

\subsection{Optimal investment strategies for different levels of \ce{H2} demand and biomass availability}
\label{subsec:optimal_design_base_scenarios}

\begin{figure}[!ht]
    \centering
    \includegraphics[width=\textwidth]{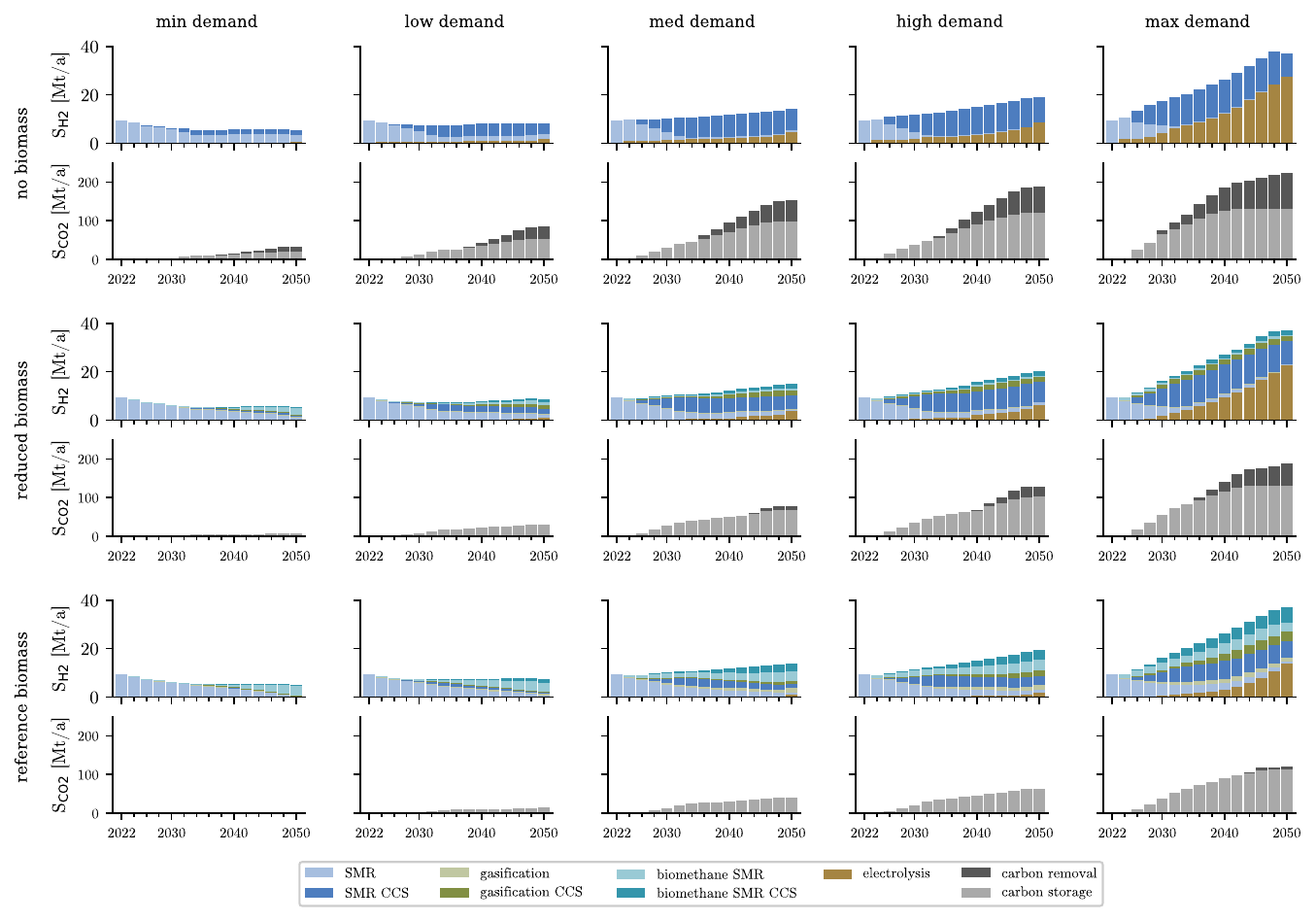} 
    \caption{Cost-optimal rollout of \ce{H2} production and \ce{CO2} capture and storage capacities ($S_{\text{\ce{H2}}}$ and $S_{\text{\ce{CO2}}}$) in Mt/a from 2022 to 2050 across the 15 design scenarios. \ce{H2} production technologies include steam methane reforming (SMR) from natural gas, biomethane reforming, biomass gasification, and water-electrolysis from electricity. SMR, biomethane reforming, and biomass gasification can be coupled with CCS. In addition, \ce{CO2} removal and \ce{CO2} storage technologies can be installed to meet the net-zero emissions target by 2050.}
    \label{fig:capacity_base_scenarios}
\end{figure}

\cref{fig:capacity_base_scenarios} visualizes the optimal rollout of the \ce{H2} production, and \ce{CO2} capture and storage capacities across all scenarios. The installed \ce{H2} production capacity depends on the expected \ce{H2} demand. In the max \ce{H2} demand scenario, the \ce{H2} production capacity for 2050 is about 7 times higher than in the min \ce{H2} demand scenario. In contrast, the \ce{H2} production technology mix and the \ce{CO2} capture and storage capacities strongly depend on the availability of biomass. 

Biomass-based \ce{H2} production is identified as the most cost-efficient low-carbon \ce{H2} production pathway. By replacing natural gas with biomethane, the SMR process emissions can be reduced by 82\,\%. Besides installing anaerobic digesters to produce biomethane, no additional investments are required, and existing SMR capacities can continue to be used. 
Furthermore, the coupling of biomass-based \ce{H2} production with CCS results in net-negative emissions, offsetting \ce{CO2} emissions that occur at other stages of the supply chain. Therefore, as a first step, scenarios with biomass availability replace natural gas feedstocks with biomethane. With increasing annual decarbonization targets, \ce{H2} production is complemented with CCS to reduce process emissions and biomass gasification is deployed. For max \ce{H2} demands, the reference biomass potentials are insufficient to fully decarbonize \ce{H2} production, and increasing shares of electrolyzers are installed. In addition, investments in \ce{CO2} removal technologies are required to offset upstream supply chain emissions from fuel supply chain, plant manufacturing and construction phases (about 6\,Mt/a). The same considerations apply for scenarios with reduced biomass availability and medium-max \ce{H2} demands, which require between 10-56\,Mt/a \ce{CO2} removal capacities by 2050 to achieve the net-zero emissions target. 

Without biomass feedstock, low-carbon \ce{H2} is produced via SMR-CCS from natural gas and electrolysis of renewable electricity, and \ce{CO2} removal technologies are installed to eliminate residual emissions and achieve the NZE target. In configurations where the \ce{H2} demand is expected to decrease or remain similar to today's values (i.e., min-low demand scenarios, \cref{fig:capacity_base_scenarios}), \ce{H2} is predominantly produced via SMR-CCS, which is associated with larger \ce{CO2} emissions, but lower cost compared to electrolysis. Electrolyzer capacities are continuously expanded in scenarios where \ce{H2} demand is expected to increase. 

Independently of the biomass availability, electrolyzers do not play a role in min-low \ce{H2} demands and are only deployed if medium-max demands are expected. Two factors contribute to the increasing deployment of electrolyzers. First, the unit cost of electrolyzers is expected to decrease by 60\,\% until 2050, making electrolyzers more cost-competitive. Second, the residual emissions of electrolyzers are 4-10 times lower compared to SMR-CCS (0.6-1.6\,ton\textsubscript{\ce{CO2}eq.}/ton\textsubscript{\ce{H2}} vs. 5.9\,ton\textsubscript{\ce{CO2}eq.}/ton\textsubscript{\ce{H2}}). Hence, offsetting the residual emissions from electrolysis requires significantly smaller capacities of \ce{CO2} removal technologies and \ce{CO2} storage.
The literature report a wide range of electrolyzer cost estimates. However, even when assuming a very optimistic cost evolution for electrolyzers, where the investment costs are 10\,\% lower in 2022 and 30\,\% lower in 2050 with respect to the reference case, electrolyzer capacities remain small and are only deployed at a larger scale in high and max \ce{H2} demand scenarios (\cref{suppsec:optimistic_electrolysis}).

The \ce{CO2} removal technologies are located near the \ce{CO2} storage sites to reduce \ce{CO2} transport costs. Land requirements for \ce{CO2} removal technologies range between 0.2-0.4\,km\textsuperscript{2} \cite{Ozkan2022,climeworks2023}. Depending on the scenario the capacity of the \ce{CO2} removal technologies range between 6 to 92\,Mt/a which translates to 1.2-18\,km\textsuperscript{2} in the best case, and 2.4-37\,km\textsuperscript{2} in the worst case. By far, the largest \ce{CO2} removal capacity is installed close to the Northern Lights \ce{CO2} storage site in Norway where about 30\,Mt/a \ce{CO2} are removed from the air each year, requiring an area of 6-12\,km\textsuperscript{2} (about 840-1,680 soccer fields). To provide more context, this corresponds to less than 1\,\% of the open free available area \cite{ssb2023}. In the remaining regions, the \ce{CO2} removal capacity is below 8.5\,Mt/a (1.7-3.4\,km\textsuperscript{2}). 

Finally, the regional \ce{CO2} storage capacity is limited by the capacity of existing and announced \ce{CO2} storage projects, which corresponds to 132\,Mt/a  \cite{IOGP2023}. The full \ce{CO2} storage potential is used by 2050 if max \ce{H2} demands are expected and less than the reference biomass potentials are available for \ce{H2} production. 

\subsection{Hydrogen and carbon dioxide transport networks}

\begin{figure}[!ht]
    \centering
    \includegraphics[width=\textwidth]{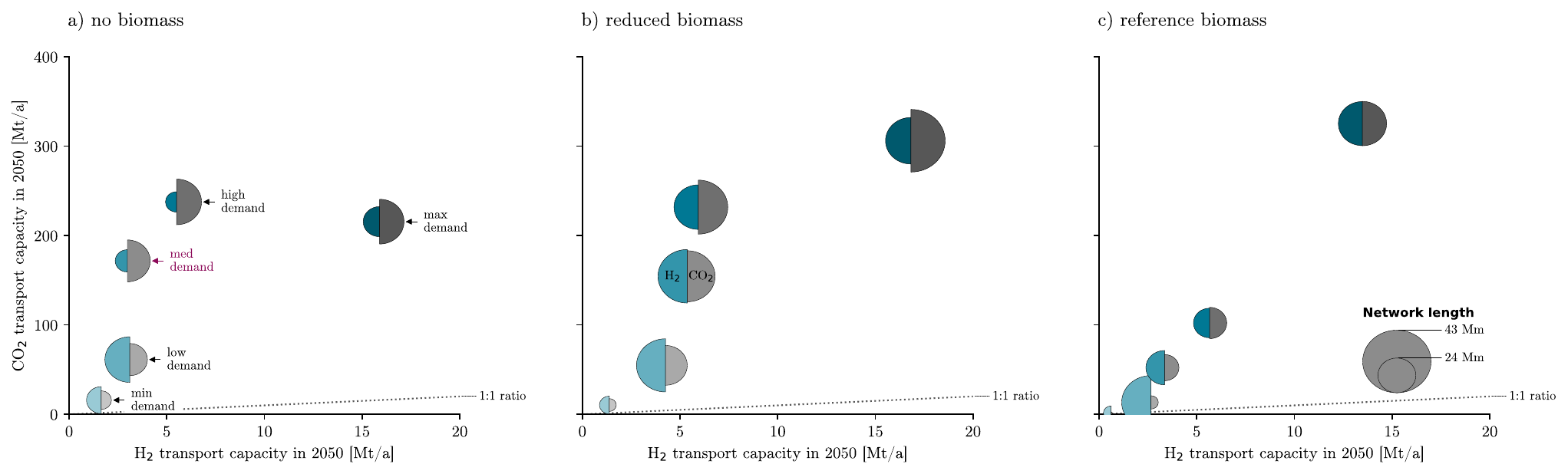}
    \caption{\ce{H2} and \ce{CO2} transport network capacity and length for the 15 design scenarios. The location of the bubbles indicates the cumulative \ce{H2} (x) and \ce{CO2} transport network capacity (y). The bubble size indicates the \ce{H2} (left, petrol color) and \ce{CO2} network length (right, grey color). The shading indicates the level of \ce{H2} demand, where lighter shades indicate lower demands, and darker shades indicate higher demands.}
    \label{fig:hydroge_carbon_network}
\end{figure}

A \ce{H2} transport network is required to decouple \ce{H2} production and demand. In addition, a \ce{CO2} transport network is needed to transport the captured \ce{CO2} to the \ce{CO2} storage sites. \cref{fig:hydroge_carbon_network} compares the \ce{H2} and \ce{CO2} transport networks in 2050 with respect to (i) the network capacity (x and y location of the bubble for \ce{H2} and \ce{CO2}, respectively), and (ii) the network extension (size of the right and left half of the bubble for \ce{H2} and \ce{CO2}, respectively). The capacity of the \ce{H2} and \ce{CO2} networks is strongly connected to the \ce{H2} demands, and in general, the transport volumes and the network capacities increase with increasing \ce{H2} demands. 

While the \ce{H2} network transport capacity is comparable for different levels of biomass, the spatial extension of the network varies greatly. System designs that rely on biomass deploy larger \ce{H2} transport infrastructure to overcome the additional spatial constraints imposed by the heterogeneous biomass availabilities and, therefore, more widespread transport infrastructures are required. To enable CCS, the \ce{H2} transport infrastructure is accompanied by \ce{CO2} transport infrastructure to transport the captured \ce{CO2} to the storage sites, and thus, require more extensive \ce{CO2} networks. While the capacity of the \ce{CO2} networks for reference and reduced biomass are comparable, in case of reduced biomass potentials, the \ce{CO2} network covers larger distances to access the available biomass feedstocks and enable the coupling with CCS. 

In contrast, system designs that do not rely on biomass produce \ce{H2} closer to the demand locations, resulting in smaller, more local \ce{H2} transport networks. \ce{H2} is largely produced via SMR-CCS from natural gas and electrolysis from renewable electricity, and \ce{CO2} removal technologies are installed to offset residual emissions from \ce{H2} production and up-stream emissions from plant manufacturing and construction (\cref{fig:capacity_base_scenarios}). In earlier years and for lower demands, \ce{H2} is predominantly produced via SMR-CCS. However, for larger \ce{H2} demands, SMR-CCS is complemented by increasing shares of electrolyzers, reducing the need for \ce{CO2} transport from the \ce{H2} production site to the \ce{CO2} storage locations (compare infrastructure size for min-low demand and medium-max demand for no biomass in~\cref{fig:hydroge_carbon_network}a). 

\subsection{Levelized cost of hydrogen for different levels of hydrogen demand and biomass availability}
\label{subsec:LCOH}

\begin{figure}[h!]
    \centering
\includegraphics[width=\textwidth]{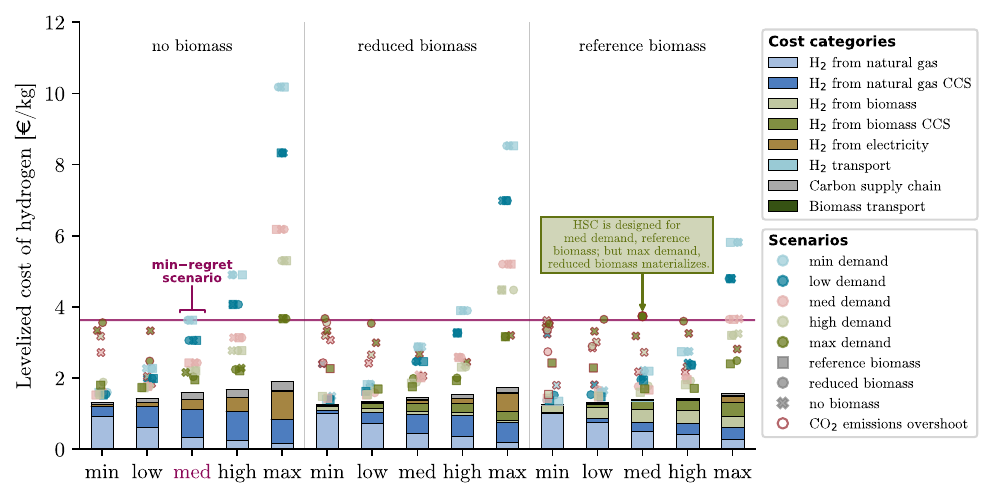}
    \caption{Levelized cost of \ce{H2} (LCOH) for the across all scenarios $s \in \mathcal{S}$. The LCOH is computed as the net present cost divided by the net present \ce{H2} production. The bars indicate the LCOH of the design scenarios. The markers above the bars indicate the LCOH that arises if the supply chain is initially designed for scenario $a \in \mathcal{S}$, but scenario $b \in \mathcal{S}$ materializes. A red marker edge color indicates infeasible scenarios where the annual \ce{CO2} emissions targets cannot always be fulfilled. }
    \label{fig:LCOH}
\end{figure}

\Cref{fig:LCOH} presents the LCOH (i) for each design scenario (i.e., the system is designed and operated on the same scenario - stacked bars) and (ii) the results of the out-of-sample approach (i.e., the system is designed for one scenario and operated on all other 14 scenarios - markers above each bar). 
The marker shape and color indicate which out-of-sample scenario the supply chain design is operated on. Furthermore, a red marker edge indicates that the out-of-sample scenario is infeasible, i.e., it is impossible to adjust the system design quickly enough such that the annual \ce{CO2} emissions targets are always satisfied. 

The lowest LCOH is observed for designs with reference biomass. Here, the LCOH is 9-13\,\% lower compared to design for reduced or no biomass, respectively. Furthermore, we observe a cost increase of up to 44\,\% with increasing \ce{H2} demand. This cost increase is attributed to larger investments in electrolyzers, and \ce{CO2} capture, transport, and storage infrastructure, which are not required in scenarios with lower \ce{H2} demands.
Assuming that a pan-European \ce{CO2} transport infrastructure may be deployed independent from the decarbonization of the hard-to-abate industries investigated here, and therefore, \ce{CO2} transport infrastructure is available at little to no cost, \ce{H2} production capacities from SMR-CCS are expanded (\cref{supp:optimistic_CCTS}). This is true, especially for scenarios with reference biomass availability \cref{fig:evolution_capacities_diff_low_CCS}. Nevertheless, the impact of an inexpensive \ce{CO2} transport infrastructure is marginal with a LCOH cost reductions below 5\,\% across scenarios.

If a different scenario materializes than the HSC was initially designed for, the HSC design has to be adjusted to the new conditions and the LCOH increases. Depending on the magnitude of those changes, it may not be possible to adapt the initial supply chain design such that the \ce{CO2} emission targets are fulfilled at all times (\cref{supp:co2_emissions}). HSC designs where the \ce{CO2} emission targets cannot be fulfilled at all times are considered infeasible. These infeasibilities occur predominantly when the \ce{H2} demand increases drastically (e.g., instead of min-low demands, high-max demands materialize) or when biomass availability is significantly overestimated during the design phase due to difficulties in scaling up the capacity of low-carbon \ce{H2} production and \ce{CO2} removal technologies. 

The speed at which technology capacities can be expanded is described by the technology expansion rate as a function of the existing technology capacity \cref{subsec:optimization_model}. The number of infeasible scenarios is lower for more conservative system designs, i.e., systems designed without biomass and for medium to max \ce{H2} demands, where larger shares of electrolyzers and \ce{CO2} removal technologies allow for a quicker scale-up of the already existing capacities. In contrast, supply chain designs that initially rely largely on biomass-based \ce{H2} production technologies are often not able to switch strategies and scale up the capacity of electrolyzers, SMR-CCS, and \ce{CO2} removal technologies quickly enough to meet the demand for low-carbon \ce{H2}.  
Higher technology expansion rates decrease the number of infeasible scenarios as they allow for a quicker expansion of the existing capacity and, thereby, offer greater flexibility to react and adapt the initial supply chain design to changes (\cref{fig:LCOH_td}). Nevertheless, the investment strategies remain robust for low and high technology expansion rates, and designing HSCs for medium \ce{H2} demands, without biomass remains the minimum-regret strategy. 

\subsection{Minimum-regret solution}
\label{subsec:min_regret_solution}

The minimum-regret solution is identified based on two criteria: (1) the solution results in the lowest LCOH in the worst scenario (min-max cost criteria), while (2) meeting the annual \ce{CO2} emissions targets (feasibility criteria). 
Overall, only three out of 15 design scenario designs meet the feasibility criteria of complying with the annual \ce{CO2} emissions target at all times and across all out-of-sample scenarios; these are the no biomass scenarios for medium to max \ce{H2} demands (\cref{fig:LCOH}). Out of these three scenarios, the designs for medium \ce{H2} demands and no biomass is identified as the minimum-regret solution as it results in the lowest LCOH in the worst case (\cref{subsec:minimum_regret_solution}). The minimum-regret solution remains robust for different technology expansion rates (\cref{fig:LCOH_td}).  
In the minimum-regret scenario, about 63\,\% of the \ce{H2} demand in 2050 is met via SMR-CCS, and about 34\,\% via electrolysis from renewable electricity. Depending on the \ce{H2} demand of the out-of-sample scenario, the electrolyzer capacity must be expanded by up to 22\,Mt/a. 
While planning for medium demands and reference biomass availability leads to similar LCOH across scenarios, the net-zero emission target for medium-max \ce{H2} demands is likely not achieved by 2050 if less biomass is available than expected. Therefore, designing HSCs for medium \ce{H2} demands and without the availability of biomass feedstocks is identified as the minimum-regret strategy (\cref{subsec:min_regret_solution}).

\begin{figure}[h!]
    \centering
    \includegraphics[width=.7\textwidth]{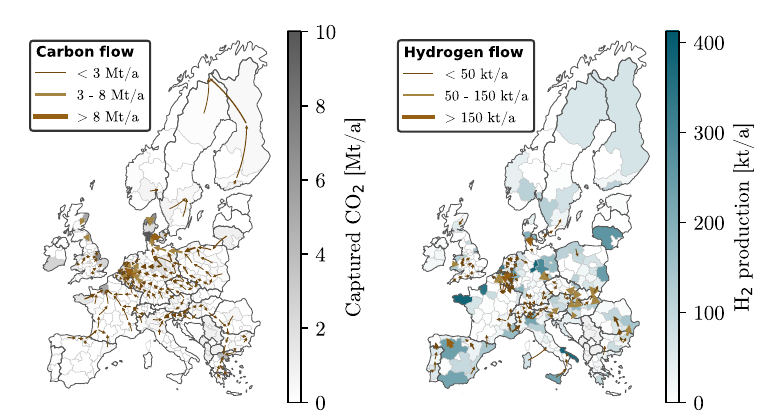}
    \caption{a) \ce{H2} production and transport, and b) \ce{CO2} capture and transport infrastructure in the minimum-regret scenario with no biomass and medium \ce{H2} demands.} 
    \label{fig:min_regret_networks}
\end{figure}

A medium-sized \ce{H2} and \ce{CO2} transport infrastructure is built (3\,Mt/a and 172\,Mt/a, respectively, \cref{fig:min_regret_networks}). 
In particular, the regions of Belgium, Netherlands, and the west of Germany are well connected by \ce{H2} and \ce{CO2} transport infrastructure. This offers enough flexibility to expand existing capacities without leading to large unused capacities in case smaller \ce{H2} demands materialize. Compared to other configurations, the mean network utilization rates remain high (\cref{fig:utilisation_rate_network}).

\section{Discussion}
\label{sec:discussion}

\subsection{The role of biomass in decarbonizing \ce{H2} production} 

Biomass-based \ce{H2} production is identified as the most cost-efficient low-carbon \ce{H2} production technology due to low \ce{H2} production cost and \ce{CO2} intensities. Especially in the initial phase of the transition, when the capital cost of electrolyzers is still high, biomass-based \ce{H2} production dominates the technology mix (\cref{fig:capacity_base_scenarios} and \cite{Ganter2023}). This finding is consistent with \cite{Lane2021a}, who investigate the optimal \ce{H2} production technology mix under cost uncertainties. In comparison to biomass-based \ce{H2} production, the LCOH from electrolysis and renewable electricity is 3-5 times higher, hindering the deployment of electrolyzers at large scales \cite{Wang2022}. Even if the projected cost reductions of 2\,\euro{}/kg\textsubscript{\ce{H2}} can be achieved by 2040 \cite{Terlouw2019}, the LCOH from electrolyzers is expected to remain high. Only SMR-CCS can achieve lower \ce{H2} production cost than biomass-based technologies (about 1.5\,\euro{}/kg\textsubscript{\ce{H2}}), at the expense of higher process emissions (+2.5-3\,kg\textsubscript{\ce{CO2eq.}}/kg\textsubscript{\ce{H2}}). Therefore, the cost-competitiveness of electrolytic \ce{H2} is viewed as an unrealistic prospect in the medium-term without appropriate policy support \cite{Staffel2019}.

However, the biomass availability is uncertain as multiple sectors compete for it. For instance, biomass can be used as a fuel in the transport sector to decarbonize aviation or heavy-duty transport \cite{Becattini2021,Daiogolou2015}, or it can provide dispatchable, flexible energy in the power sector \cite{Bogdanov2019}. While the LCOH is lower in scenarios that rely on biomass in their decarbonization strategy, planning without biomass leads to more robust solutions. \cref{fig:annual_emissions} visualizes the annual \ce{CO2} emissions for each design scenario. When planning without biomass, most scenarios can eventually achieve a net-zero supply chain design. Only when significantly higher \ce{H2} demands materialize than planned (see configurations for min-low demand), the annual \ce{CO2} emission targets are overshot during the transition. 
This observation is robust across the investigated range of technology expansion rates \cref{supp:sensitivity_td}. 

If biomass is included in the technology mix, policymakers must ensure that biomass will be dedicated to low-carbon \ce{H2} production. Otherwise, it might not be possible to adapt the decarbonization strategy and scale up alternative \ce{H2} production and \ce{CO2} removal technologies quickly enough to satisfy the \ce{H2} demands in compliance with stricter \ce{CO2} emission limits (\cref{fig:LCOH}). 

Currently, biomass is predominantly used for heating and cooling (about 75\,\% in 2018) \cite{BiomassEU2024}. However, existing studies indicate that using biomass to decarbonize transport and industry is more attractive than using biomass as a dispatchable, flexible energy source for electricity production \cite{Wu2023}. The current bioeconomy strategy of the EU focuses on increasing the sustainability and circularity, but does not provide clear guidelines on the use of biomass \cite{EUComissionBioenergy2018}. Therefore, we recommend extending existing EU and national bioeconomy strategies to provide clear guidelines on the strategic use of the limited biomass resources; steering the use of biomass to sectors where biomass provides a cost-efficient decarbonization option and which lack alternatives.

\subsection{The role of carbon capture, transport, and storage}

\ce{CO2} capture, transport, and storage (CCTS) infrastructure plays a central role in decarbonizing \ce{H2} production. More specifically, when coupling \ce{H2} production with CCS, large amounts of \ce{CO2} are captured at the production site, transported, and stored underground. When \ce{H2} production is based on electrolysis, \ce{CO2} removal technologies are installed to remove residual and upstream \ce{CO2} emissions. The \ce{CO2} removal technologies are typically located close to the \ce{CO2} storage sites to reduce transport costs, and thus, require less extensive \ce{CO2} transport infrastructure. 

For scenarios with large \ce{H2} demands and reduced or no biomass, existing and planned \ce{CO2} storage sites are fully utilized to achieve net-zero \ce{H2} production. A reduced \ce{CO2} storage availability would lead to an increased installation of electrolysers. Nevertheless, a minimum of about 7, 11, and 50\,Mt\textsubscript{\ce{CO2}}/a \ce{CO2} removal and storage capacity would be required to offset upstream emissions and realize net-zero \ce{H2} production for minimum, medium, and maximum demand, respectively.

The need for CCTS infrastructure is widely recognized \cite{Bui2018}, and according to the International Association of Oil \& Gas Producers (IOGP) \cite{IOGP2022}, Europe requires a minimum of 0.5-1\,Gton\textsubscript{\ce{CO2}}/a of \ce{CO2} storage by 2050 to reach its targets. However, the capacity of existing and currently planned \ce{CO2} storage projects in Europe only corresponds to 13-26\,\% of the storage capacity envisioned by \cite{IOGP2022}. For reference, the European hard-to-abate industry alone currently emits between 0.7-1\,Gton\textsubscript{\ce{CO2}}/a. Similar observations are made on a global scale, where only 10\,\% of the required \ce{CO2} storage capacity might be available by 2050, considering the historically observed expansion rates for \ce{CO2} storage \cite{Martin-Roberts2021}. This suggests that rapid changes and coordinated efforts are required to enable CCS across sectors and foster the energy transition. 

\subsection{Implications of a constrained technology expansion}

Technology expansion rates are computed based on historically observed annual growth rates, as they do not appear to change significantly over time \cite{Wilson2009,Leibowicz2016,Mannhardt2023b}. Previous studies indicate that energy systems models tend to overestimate the speed at which emerging technologies can penetrate the market \cite{Leibowicz2016,Bertram2015}. Technology expansion constraints can be used to limit the speed at which the capacity of a technology can be expanded (\cref{subsec:optimization_model}). 

Results indicate that the feasibility and cost of out-of-sample designs strongly depend on the availability of \ce{CO2} removal technologies. The reduced investments in \ce{CO2} removal technologies in design scenarios that rely on biomass and the coupling with CCS pose challenges when biomass does not become available as planned, especially when larger \ce{H2} demands materialize than initially planned for. The results further indicate that planning ahead and investing in sufficient low-carbon \ce{H2} production capacities (about 9.6\,Mt/a by 2030) is necessary to scale-up low-carbon \ce{H2} production and facilitate larger low-carbon \ce{H2} demands. Otherwise, the transition might be delayed and the decarbonization targets cannot be met (\cref{fig:LCOH} and \cref{fig:annual_emissions}). The planning horizon of 2022-2030 is especially important. Both, the design scenarios for no biomass and low and mean demands fail to achieve the annual emission targets in 2024-2030 because insufficient low-carbon \ce{H2} production capacities delay the scale-up of SMR-CCS and electrolyzer capacities, and instead, larger shares of the existing, natural-gas based SMR have to be deployed to meet the increased \ce{H2} demands. \cite{Shi2021} show that large demands for low-carbon \ce{H2} (e.g., induced by high carbon taxation) can stimulate the adoption of low-carbon technologies. Hence, developing a stringent European \ce{H2} strategy could promote investments in low-carbon \ce{H2} infrastructure and foster its widespread use. 
The lack of a dedicated \ce{H2} and \ce{CO2} transport infrastructure \cite{Schlund2022,InternationalEnergyAgencyIEA2020a}, as well as uncertainties around material selection and process operation \cite{Bui2018} hinder investments in low-carbon \ce{H2} and CCTS infrastructure. While the creation of niche markets for low-carbon \ce{H2} could accelerate its adoption, this also necessitates a coordinated scale-up of the required low-carbon \ce{H2} and CCTS transport infrastructure \cite{Schlund2022}.

\section{Conclusion}
\label{sec:conclusion}

This work investigates the optimal design and rollout of \ce{H2} supply chains (HSCs) to achieve a net-zero European hard-to-abate industry. We determine the optimal technology portfolio of net-zero \ce{H2} production technologies while accounting for uncertainties in the evolution of \ce{H2} demand and biomass availability. We consider a multi-year time horizon from 2022 to 2050 and include several \ce{H2} production pathways, namely water-electrolysis from renewable electricity, SMR from natural gas and biomethane, and biomass gasification. \ce{H2} production from biomass and natural gas can be coupled with CCS. In addition, \ce{CO2} removal technologies can be installed to remove residual \ce{CO2} emissions. Finally, \ce{CO2} and biomass transport infrastructures are designed alongside the HSC.

We follow a scenario-based uncertainty quantification approach and define 15 scenarios, considering five levels of \ce{H2} demand and three levels of biomass availability. We determine the optimal HSC design for the individual scenarios (design scenarios) and evaluate their performance in case alternative scenarios materialize (out-of-sample approach). Additional investments can be made to adapt the supply chain designs to the new operating conditions. However, these additional investments are limited by technology expansion constraints. The solutions are evaluated and compared based on the levelized cost of \ce{H2}, and a minimum-regret design is determined based on two criteria: (i) the ability to meet the annual \ce{CO2} emissions target and achieve net-zero emissions by 2050, and (ii) the lowest levelized cost of \ce{H2} in case the worst-case scenario materializes.

\ce{H2} production via SMR-CCS of natural gas and biomethane reforming are the most cost-efficient low-carbon \ce{H2} production pathways. Particularly attractive is the coupling of biomass-based \ce{H2} production with CCS, a process which achieves a \ce{CO2} removal, and can thereby offset emissions at other stages in the supply chain. Investments in electrolyzers are delayed due to high initial capital costs. 

To harness the full potential of regionally-constrained biomass and to decouple \ce{H2} production and demands, \ce{H2} transport infrastructure is required. Furthermore, \ce{CO2} capture, transport, and storage infrastructure (CCTS) plays a central role in a low-carbon \ce{H2} economy. Especially for larger \ce{H2} demands, a pan-European CCTS infrastructure is required to mitigate emissions from \ce{H2} production and enable CCS. A minimum \ce{CO2} storage capacity of 2.7-50\,Mt\textsubscript{\ce{CO2}}/a is needed. Hence, a coordinated effort on a European level is required to accelerate the development and deployment of CCTS. 

In general, the LCOH is 9\,\%-13\,\% lower for configurations that include biomass in their decarbonization strategy. However, a strategy relying on biomass is not robust if biomass availability is reduced. In this case, it is often impossible to adapt the system design and scale up alternative low-carbon \ce{H2} production technologies to meet the \ce{H2} demand and the decarbonization targets. In contrast, planning without biomass leads to more robust supply chains. 

We also find that planning for medium \ce{H2} demands and installing a sufficient \ce{H2} production capacity (about 9.6\,Mt/a by 2030) is required to facilitate the scale-up of existing low-carbon \ce{H2} production capacities to accommodate larger \ce{H2} demands. Otherwise, the scale-up of the low-carbon \ce{H2} production capacity might be delayed, ultimately resulting in a failure to achieve the annual \ce{CO2} emissions targets, even if technology expansion rates are high.

A comprehensive European \ce{H2} and bioenergy strategy could help navigate these large uncertainties by delineating clear directives on (i) the envisioned applications of \ce{H2} and (ii) the role of biomass in the energy transition.
Dedicating further research on the role of low-carbon \ce{H2} in decarbonizing the selected hard-to-abate industries and identifying no-regret applications can reduce uncertainty regarding future levels of low-carbon \ce{H2} demand and guide policymakers. However, if the large uncertainty surrounding the future demand for low-carbon \ce{H2} pertains, this may further delay the scale-up of low-carbon HSCs. 

Overlooking the importance of CCTS infrastructure can hinder the transition and jeopardize the realization of climate goals. \ce{H2} and \ce{CO2} transport infrastructure is required independently of the low-carbon \ce{H2} demand. Fostering investments in regional \ce{H2} and \ce{CO2} transport infrastructure that can be expanded if larger low-carbon \ce{H2} demands materialize can support a successful transition to net-zero. Incentivizing investments in low-carbon \ce{H2} production capacities can further reduce the risks first-movers face (slow market dynamics, high low-carbon \ce{H2} production costs, lack of transport infrastructure) and facilitate the scale-up of low-carbon \ce{H2} markets. 

Finally, although biomass-based \ce{H2} production is lower in cost, including biomass in the decarbonization strategy can impose risks if these biomass feedstocks are not made available. Therefore, planning without biomass feedstocks is identified as the minimum-regret strategy. Extending the EU bioenergy strategy and providing clear directives on the sectors that will get access to the limited biomass potentials can guide investments in the low-carbon \ce{H2} production technologies. Here, additional research can support policymakers in identifying the sectors where biomass plays a key role in the transition to net zero.

\section*{Code and data availability}
The code and input data to reproduce the results presented in
this work is available on Zenodo: \url{https://zenodo.org/doi/10.5281/zenodo.10930801}.

\section*{Acknowledgments}
This work is supported by the Swiss National Science Foundation, grant no. 200021-182529, and the Swiss Federal Office of Energy as part of the SWEET PATHFNDR project.

\newpage
\clearpage

\renewcommand{\appendixname}{Supporting Information}

\appendix

\renewcommand{\thesection}{Section S\arabic{section}}
\section*{Supporting Information}
\renewcommand{\thefigure}{S\arabic{figure}}
\setcounter{figure}{0}
\renewcommand{\thetable}{S\arabic{table}}
\setcounter{table}{0}

\section{Input data to the optimization problem}
\label{supp:input_data}

\subsection{Energy carrier prices and carbon intensities}
\label{supp:input_data_feedstocks}

The country-specific grid electricity prices and \ce{CO2} intensities are taken from \cite{EUreference2020}. The country-specific natural gas prices are taken from \cite{EurostatNaturalGasPrice}. The yearly evolution of the natural gas prices is estimated following \cite{EnergyMarketUpdate2023}. The regional wet and dry biomass prices are taken from \cite{Ruiz2019}. Finally, the \ce{CO2} intensity of natural gas, wet biomass, and dry biomass are taken from  \cite{Antonini2020,Antonini2021}. 


\subsection{Conversion and transport technologies}
\label{supp:input_data_technologies}
The parameters describing the technology cost of the conversion technologies are described in \cref{tab:capex_conver_techs,tab:economic_parameters_techs}. The parameters describing the performance of the conversion technologies are reported in \cref{tab:technological_parameters_production_techs,tab:techno_economic_parameters_transport_techs}. The capacity factors and capacity limits for renewable energy technologies are taken from \cite{Gonzalez-Aparicio2021}. Finally, the techno-economic parameters of the transport technologies are described in \cref{tab:techno_economic_parameters_transport_techs}. The transport distances for the transport technologies are approximated by the Haversine distance between the centroids of two neighboring regions.

\begin{table}[htb]
\small
\centering
\caption{Capital investment costs for the available conversion technologies. bm = biomethane; e = electricity.}
\label{tab:capex_conver_techs}
\begin{tabular}{lllllllll}
\toprule
\textbf{Technologies} & unit & 2022 & 2025 & 2030 & 2035 & 2040 & 2045 & 2050 \\
\midrule
Electrolysis \cite{IEA2019,Victoria2022} & \euro/kW\textsubscript{\ce{H2}} &  1079 & 913 & 747 & 622 & 498 & 456 & 415 \\
SMR  \cite{IEA2019} & \euro/kW\textsubscript{\ce{H2}} & 840 & 824 & 809 & 793 & 777 & 761 & 745 \\
SMR-CCS  \cite{IEA2020Assumptions} & \euro/kW\textsubscript{\ce{H2}} & 1551 & 1404 & 1256 & 1237 & 1219 & 1200 & 1182 \\
Gasification \cite{Kuba2018,Terlouw2019} & \euro/kW\textsubscript{\ce{H2}} & 1327 & 1287 & 1247 & 1207 & 1167 & 1128 & 1088 \\
Gasification-CCS \cite{Kuba2018,Terlouw2019} & \euro/kW\textsubscript{\ce{H2}} & 2449 & 2216 & 1983 & 1953 & 1924 & 1895 & 1866\\
Anaerobic digestion \cite{Terlouw2019,Victoria2022}  & \euro/kW\textsubscript{bm} & 1224 & 1164 & 1103 & 1074 & 1048 & 1019 & 993\\
Wind onshore \cite{TYNDP2021} & \euro/kW\textsubscript{e} & 1353 & 1303 & 1252 & 1217 & 1182 & 1174 & 1165\\
Wind offshore \cite{TYNDP2021} & \euro/kW\textsubscript{e} & 2115 & 2009 & 1903 & 1827 & 1751 & 1732 & 1712\\
Solar-PV (rooftop) \cite{TYNDP2021} & \euro/kW\textsubscript{e} & 1364 & 1156 & 949 & 875 & 800 & 726 & 652\\
Solar-PV (utility-scale) \cite{TYNDP2021} & \euro/kW\textsubscript{e} & 640 & 547 & 455 & 426 & 398 & 381 & 364\\
\ce{CO2} removal \cite{Fasihi2019} & \euro/(kg\textsubscript{\ce{CO2}}/h) & 7139 & 5225 & 3311 & 2816 & 2321 & 2133 & 1945\\
\ce{CO2} storage \cite{ZEP2011} &  \euro/(kg\textsubscript{\ce{CO2}}/h) & 210 & 210 & 210 & 210 & 210 & 210 & 210 \\
\bottomrule
\end{tabular}
\end{table}

\begin{table}[htb]
\small
\centering
\caption{Fixed and variable operation and maintenance (0\&M) costs, technology lifetime, and construction time (build time) for the available conversion technologies. bm = biomethane; e = electricity.}
\label{tab:economic_parameters_techs}
\begin{tabular}{lllll}
\toprule
\textbf{Technologies} & \textbf{fix O\&M} [\%] & \textbf{var O\&M} [\euro/kW\textsubscript{\ce{H2}}] & \textbf{Lifetime} [years] & \textbf{Build time} [years]\\
\midrule
Electrolysis \cite{IEA2019}  & 1.5 & 0  & 10 & 2\\
SMR  \cite{IEA2019} & 4.7 & 0  & 25 & 2 \\
SMR-CCS  \cite{IEA2020Assumptions}  & 3 & 0  & 25 & 2\\
Gasification \cite{Kuba2018,Terlouw2019}  & 5 & 0  & 20 & 2\\
Gasification-CCS \cite{Kuba2018,Terlouw2019}  & 5 & 0  & 20 & 2\\
Anaerobic digestion \cite{Terlouw2019,Victoria2022}  & 6 & 0 \euro/kW\textsubscript{bm} & 20 & 2\\
Wind onshore \cite{TYNDP2021}  & 1 & 14\euro/kW\textsubscript{e} & 30 & 2 \\
Wind offshore \cite{TYNDP2021}  & 2 & 39\euro/kW\textsubscript{e}  & 30 & 2\\
Solar-PV (rooftop) \cite{TYNDP2021}  & 2 & 12\euro/kW\textsubscript{e} & 40 & 2\\
Solar-PV (utility-scale) \cite{TYNDP2021}  & 2 & 8 \euro/kW\textsubscript{e} & 40 & 2\\
\ce{CO2} removal \cite{Fasihi2019} & 4 & 0 \euro/t\textsubscript{\ce{CO2}} & 25 & 2\\
\ce{CO2} storage \cite{ZEP2011} & 6 & 0 \euro/t\textsubscript{\ce{CO2}} & 40 & 2\\
\bottomrule
\end{tabular}
\end{table}

\begin{table}[htbp]
\small
\centering
\caption{Technological parameters of the available production technologies. If the electricity balance is negative, electricity has to be provided to the system. g = natural gas or biomethane; e = electricity; B = dry biomass; b = wet biomass; bm = biomethane; \ce{LC} = liquid \ce{CO2}.}
\label{tab:technological_parameters_production_techs}
\begin{tabular}{lllll}
\toprule
\textbf{Production technologies} &  \textbf{Conversion} & \textbf{Efficiency} & \textbf{\ce{CO2} capture} & \textbf{\ce{CO2} intensity}\\
\midrule
Electrolysis \cite{IEA2020Assumptions} & e$\rightarrow$\ce{H2} & 0.64 & - & -\\
\arrayrulecolor{black!30}\midrule
\multirow{2}{*}{SMR \cite{Antonini2020}} & g $\rightarrow$ \ce{H2}   & 0.77 & \multirow{2}{*}{-} & \multirow{2}{*}{-}\\
                                        & \ce{H2} $\rightarrow$ \text{e} & 0.041 & &\\
\midrule
\multirow{2}{*}{SMR-CCS  \cite{Antonini2020}} &  g $ \rightarrow$ \ce{H2}      & 0.77   & \multirow{2}{*}{\SI{90}{\percent}} & \multirow{2}{*}{-}\\
                                            & \ce{H2} $\rightarrow$ \text{e} & 0.016 & &\\
\midrule
\multirow{2}{*}{Gasification \cite{Antonini2021}}  & B $\rightarrow$ \ce{H2} & 0.62    & \multirow{2}{*}{-} & \multirow{2}{*}{-}\\
                                         & \ce{H2} $\rightarrow$ \text{e}    & -0.093 & &\\
\midrule
\multirow{2}{*}{Gasification-CCS \cite{Antonini2021}} &  B $\rightarrow$ \ce{H2} & 0.62 & \multirow{2}{*}{\SI{57}{\percent}} & \multirow{2}{*}{-}\\
                                           & \ce{H2} $\rightarrow$ \text{e}  & -0.153 & &\\
\midrule
Anaerobic digestion \cite{Antonini2021} & b $\rightarrow$ \text{bm} & \SI{0.435}{\tonne\per{\mega\watt\hour}} & - & -\\
\midrule
\multirow{2}{*}{\ce{CO2} capture (retrofit) \cite{Antonini2020}} &  \ce{H2} $ \rightarrow$ \ce{H2}      &  1  & \multirow{2}{*}{\SI{90}{\percent}} & \multirow{2}{*}{-}\\
                                            & \ce{e} $\rightarrow$ \text{H2} & 0.025 & &\\
\midrule
\multirow{2}{*}{\ce{CO2} storage \cite{Becattini2022}} & \ce{LC} $\rightarrow$ \ce{LC} & 1 & \multirow{2}{*}{-} & \multirow{2}{*}{-1} \\ 
                                                          &  \ce{e} $\rightarrow$ 
                                                          \ce{LC} & \SI{38}{\kilo\watt\hour\per{\tonne}} &  \\ 
\arrayrulecolor{black!}\bottomrule
\end{tabular}
\end{table}

\begin{table}[htbp]
\small
\centering
\caption{Techno-economic parameters of the available transport technologies. \ce{GH2} = gaseous \ce{H2}; \ce{LH2} = liquid \ce{H2}; \ce{LC} = liquid \ce{CO2}, B = dry biomass; $l$ = transport distance.}
\label{tab:techno_economic_parameters_transport_techs}
\begin{tabular}{llllllll} 
\toprule
\textbf{Transport technologies} & \textbf{Carrier} & \textbf{Investment} & \textbf{O\&M} & \textbf{Operation} & \textbf{Lifetime} & \textbf{Build time} & \textbf{\ce{CO2} intensity}\\
                        & & [\euro/kW] & [\%]& [\euro/(kW km)] & [year] & [year] & [g/(kWh km)]\\
\midrule
\ce{H2} truck (gas) \cite{IEA2020Assumptions} & \ce{GH2} & 35 & 4 & 5$\cdot 10^{-5}$ & 12 & 1 & 5$\cdot 10^{-2}$\\
\ce{H2} truck (liquid) \cite{IEA2020Assumptions} & \ce{LH2} & 8 & 4 & $8 \cdot 10^{-6}$ & 12 & 1 & 8$\cdot 10^{-3}$\\
\ce{H2} pipeline \cite{IEA2020Assumptions} & \ce{GH2} & 3 & 4 & 0 & 40 & 4 & 0\\
Dry biomass truck \cite{Schnorf2021}    & B & 6 & 4 & 3$\cdot 10^{-4}$ & 12 & 1 & 8$\cdot 10^{-3}$\\
\ce{CO2} truck (liquid) \cite{Becattini2022}            & \ce{LC} & (44 + 0.04 $l$) \euro/(t/h km) & 4 & 0.5 \euro/(t km) & 12 & 1 & 7$\cdot 10^{-5}$ \si{1\per\kilo\meter}\\
\ce{CO2} pipeline \cite{Becattini2022} & \ce{LC} & 3000 \euro/(t/h km) & 1 & 0 \euro/(t km) & 45 & 4 & 1.6$\cdot 10^{-6}$ 1/km\\
\bottomrule		
\end{tabular}
\end{table}

\FloatBarrier


\subsection{Hydrogen and carbon conditioning technologies}
\label{supp:conditioning}

Due to its low energy density at ambient temperatures and pressures (2.4\,kg/m\textsuperscript{3} at 25°C and 30\,bar), \ce{H2} is typically transported as a compressed gas at ambient temperatures and high pressures (25°C, 200-350\,bar), or as a liquid (-253°C, 1\,bar)\cite{Gabrielli2020,Krasae-in2010,Reuss2017}, requiring the installation of conditioning technologies in the form of \ce{H2} compression, liquefaction, and evaporation technologies. Similar considerations apply to \ce{CO2}, which is therefore typically transported in its liquid form \cite{Becattini2022}. The conditioning technologies are modeled following \cite{Gabrielli2020} and \cite{Reuss2017}. The techno-economic parameters of the \ce{H2} conditioning technologies are reported in \cite{Ganter2024}.

\FloatBarrier

\newpage

\section{Regional hydrogen demand uncertainty}
\label{supp:regional_hydrogen_demand_uncertainty}


The regional \ce{H2} demands for ammonia production, methanol production and refineries are estimated based on a regionally resolved dataset obtained from \cite{Fuelcellsandhydrogenobservatory2021}. The regional \ce{H2} demands for steel and cement industry are estimated using the annual emissions dataset published by the European Environmental Agency (EEA). The dataset includes the location and \ce{CO2} emissions of European hard-to-abate industrial facilities emitting more than 0.1\,Mt of \ce{CO2} per year \cite{EEA2022}. The emissions data for 2018 is used since the datasets for later years are incomplete. Based on the emissions data for 2018, we can derive the industry-specific \ce{CO2} emissions per NUTS2 region in Europe. Assuming that industry-specific emissions are proportional to the production capacity, we can then estimate the regional \ce{H2} demand.  

\begin{figure}[!htp]
	\centering
	{\noindent\includegraphics[width=0.9\textwidth]{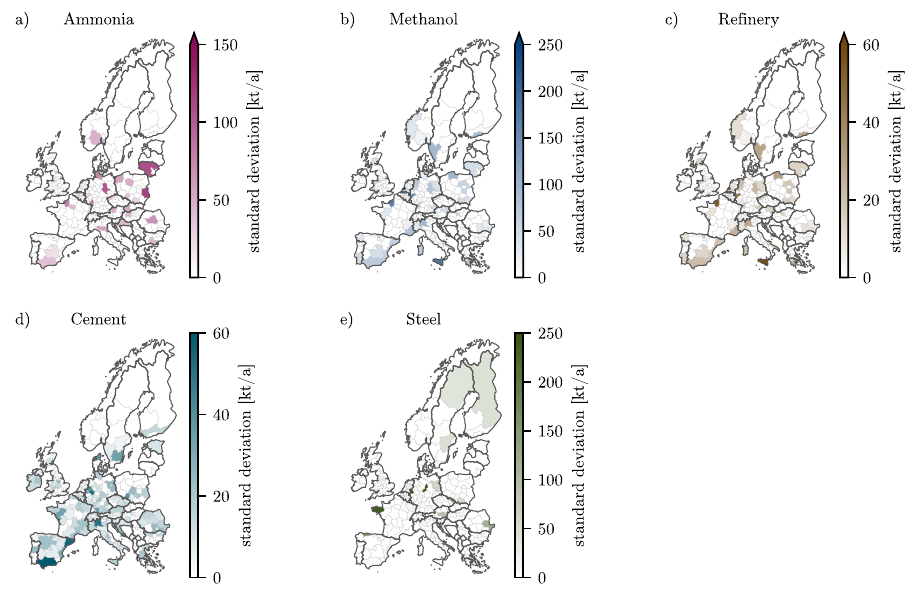}}
	\setlength{\abovecaptionskip}{5pt}
	\caption{Regional standard deviation of the \ce{H2} demand estimates for a) ammonia production, b) methanol production, c) refinery, d) cement production, and e) steel production.}
	\label{fig:map_demand_regional_deviation_sectors}
\end{figure}

We analyze the variability in the regional \ce{H2} demand estimates by comparing their standard deviation. \cref{fig:map_demand_regional_deviation_sectors} visualizes the standard deviation for each region and industry in 2050, namely (a) ammonia production, (b) methanol production, (c) refineries, (d) cement production, and (e) steel production. The standard deviation is highest for steel and methanol production, reaching up to 250\,kt/a in selected regions in the north-east of Germany and the north-west of France, and up to 280\,kt/a in the west of the Netherlands, respectively. The standard deviation for ammonia production is substantially lower, exceeding 100\,kt/a only in few select regions in the Netherlands, Germany, Poland and Lithuania. Cement and refineries exhibit the lowest standard deviations. The \ce{H2} demands from refineries are expected to decline in future years due to reduced fossil fuel demands. Cement facilities are distributed across the NUTS2 regions and similar in size, resulting in standard deviations around 20\,kt/a in most regions. 

\FloatBarrier
\newpage

\section{Technology expansion constraints}
\label{supp:technology_expansion}

The capacity of technology $h\in\mathcal{H}$ at position $p\in\mathcal{P}$ and time $t$ is expressed by $S_{h,p,t}$. Conversion technologies are installed at nodes $p=n\in\mathcal{N}$ and transport technologies can be installed at edges $p=e\in\mathcal{E}$. In each year, the technology capacity can be expanded by $\Delta S_{h,p,t}$. Existing technology capacities that are within their lifetime $l_h$ are expressed by the parameter $s^\mathrm{ex}_{h,p,\Bar{t}}$:

\begin{equation}\label{eq:technology_capacity}
S_{h,p,t} =  \underbrace{\sum^t_{\Tilde{t}=\mathrm{max}\left(0,t-l_{h}{}+1\right)} \Delta S_{h,p,\Tilde{t}}}_\text{Capacity increase}  + \underbrace{\sum^0_{\Bar{t}=min(t-l_h+1,0)} s^\mathrm{ex}_{h,p,\Bar{t}}}_\text{Capacity of existing technologies} ,
\end{equation}

Furthermore, the technology capacity $S_{h,p,t}$ and the capacity expansion $\Delta S_{h,p,t}$ are constrained by $s^\mathrm{max}_{h}{}$, and $\Delta s^\mathrm{max}_{h}{}$, respectively:
\begin{align}
\label{eq:max_technology_capacity}
&0 \leq S_{h,p,t} \leq  s^\mathrm{max}_h\\
\label{eq:max_technology_addition}
&0 \leq \Delta S_{h,p,t} \leq\Delta s^\mathrm{max}_h
\end{align}

In the out-of-sample approach, the technology capacity expansion  is additionally constrained by the technology expansion constraints. These additional constraints are modeled and parameterized following \cite{Mannhardt2023b,Leibowicz2016}. The technology expansion is limited by the technology expansion rate $\vartheta_h$ multiplied by the existing knowledge $K_{h,p,t}$, which represents the expertise and knowledge of the industry (\cref{eq:existing_knowledge}). In addition, spillover effects from one country $m$ to another $\tilde{\mathcal{M}} = \mathcal{M}\setminus\{m\}$ are considered, assuming a knowledge spillover rate $\omega=0.07$. The unbounded market share $\xi$ and the unbounded capacity addition $\zeta_h$ allow entry into niche markets \cite{Mannhardt2023b}:

\begin{align}
    0 \leq \Delta S_{h,n,t}\leq &\left(1+\vartheta_h\right)\left(K_{h,m,t}+\omega\sum_{\tilde{m}\in\tilde{\mathcal{M}}}K_{h,\tilde{m},t}\right) +\xi\sum_{\tilde{h}\in\tilde{\mathcal{I}}}S_{\tilde{h},n,t} + \zeta_h,
\end{align}

Spillover effects are not included for transport technologies $h\in\mathcal{J}\subset\mathcal{H}$, which connect regions across all nodes:
\begin{align}
    0 \leq \Delta S_{h,e,y}\leq \vartheta_h K_{h,e,t} +\xi\sum_{\tilde{h}\in\tilde{\mathcal{J}}}S_{\tilde{h},e,t} + \zeta_h.
\end{align}

To avoid unrealistically high spillover effects, the cumulative capacity additions are constrained by the cumulative existing knowledge:
\begin{align} 
\label{eq:spillover_limit}
    \sum_{p\in\mathcal{P}}\Delta S_{h,p,y}\leq \sum_{p\in\mathcal{P}}\left(\vartheta_h K_{h,p,t} +\xi\sum_{\tilde{h}\in\tilde{\mathcal{H}}}S_{\tilde{h},p,y} + \zeta_h\right).
\end{align}

where the existing knowledge $K_{h,p,t}$ is approximated by the previous capacity additions $\Delta S_{h,p,y}$ and $\Delta s^\mathrm{ex}_{h,p,y}$, and depreciated over time with the knowledge depreciation rate $\delta=0.1$:
\begin{align} \label{eq:existing_knowledge}
    K_{h,p,y} = &\sum_{\tilde{t}=t_0}^{t-1}\left(1-\delta\right)^{(t-\tilde{t})}\Delta S_{h,p,\tilde{t}} +\sum_{\hat{t}=-\infty}^{\psi(t_0)}\left(1-\delta\right)^{\left(t + (\psi(t_0)-\hat{t})\right)}s^\mathrm{ex}_{h,p,\Bar{t}}.  
\end{align}

\section{Sensitivity of the technology expansion constraint parameters}
\label{supp:sensitivity_td}

\begin{figure}[!htp]
	\centering
	{\noindent\includegraphics[width=\textwidth]{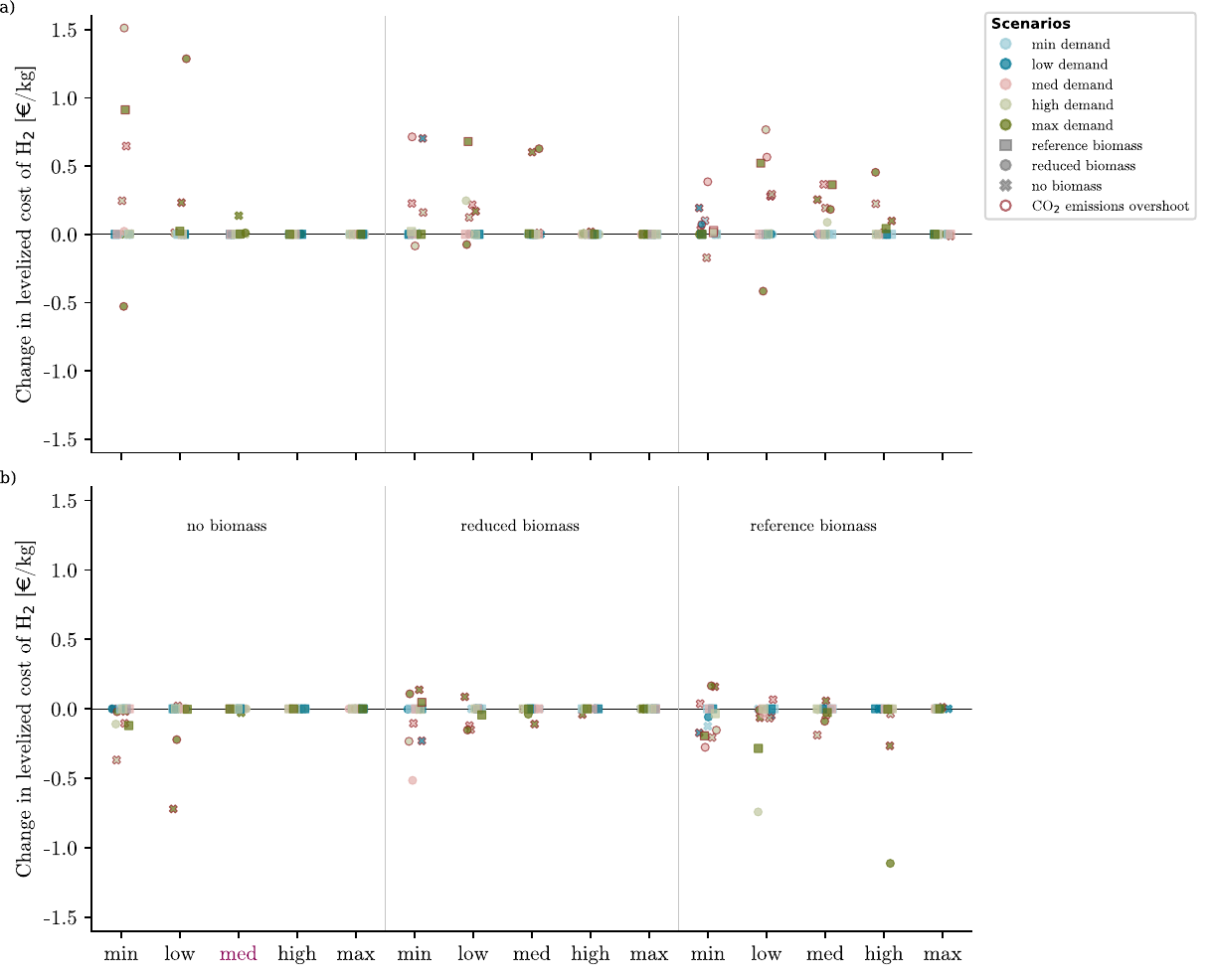}}
	\setlength{\abovecaptionskip}{-15pt}
	\caption{Change in levelized cost of \ce{H2} (LCOH) for a) low (10\,\%) and b) high (29\,\%) technology expansion rates across all scenarios $s \in \mathcal{S}$ with respect to the reference case (20\,\%). The LCOH is computed as the net present cost divided by the net present \ce{H2} production. The markers indicate the change in the LCOH with respect to the reference case presented in Fig. 6 that arises if the supply chain is initially designed for design scenario $a \in \mathcal{S}$, but out-of-sample scenario $b \in \mathcal{S}$ materializes. A red marker edge color indicates scenarios that are considered infeasible, as the \ce{CO2} emissions target cannot be fulfilled at all times. The minimum-regret design is highlighted in purple.}
	\label{fig:LCOH_td}
\end{figure}

The technology expansion constraint limits the maximum annual growth rate of a technology, and is determined by the technology expansion rate and the existing capacity of a technology. Here, we investigate the impact of different expansion rates on the levelized cost of \ce{H2} and compare the results for low expansion rates of 10\,\%, and high expansion rates of 29\,\% to the reference case of 20\,\%. \cref{fig:LCOH_td} visualize the change in the levelized cost of \ce{H2} (LCOH) for low and high expansion rates with respect to the reference case (Figure 6). A reduction in the technology expansion rate increases the systems inertia making it more difficult to quickly adapt the investment strategy and system cost increase (up to 1.5\,\euro/kg). In contrast, an increase in the technology expansion rate reduces the system inertia, allowing for a quicker implementation of changes in the investment strategy, and a reduction in cost (up to 1.1\,\euro/kg). In general, smaller systems are more sensitive to changes in the expansion rate (e.g., systems designed for min or low \ce{H2} demand), whereas the impact of the expansion rate reduces for larger systems, where the maximum annual growth rates remains high due to the larger existing capacities in the system.  

\FloatBarrier
\newpage

\section{Network utilization}
\label{supp:network_utilisation} 

\begin{figure}[h!]
    \centering
    \includegraphics[width=.7\textwidth]{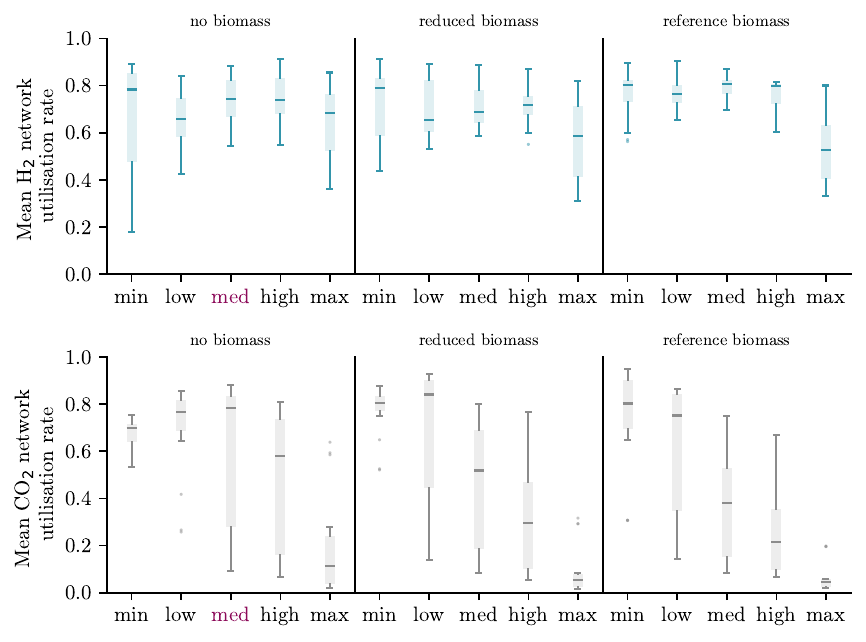}
    \caption{Boxplots of the mean network utilization rate in 2050 for the \ce{H2} and \ce{CO2} transport infrastructures for each design scenario. The minimum-regret design is highlighted in purple.} 
    \label{fig:utilisation_rate_network}
\end{figure}

\cref{fig:utilisation_rate_network} shows the utilization rate of the \ce{H2} and \ce{CO2} transport networks in 2050 across the 15 design scenarios. The network utilization rate in 2050 is computed as the carrier flow divided by the available network capacity, and the variability within each design scenario stems from the 14 out-of-sample scenarios. 
In general, the mean network utilization rates are lower for higher \ce{H2} demands. Supply chains that are initially designed for low \ce{H2} demands are characterized by small, local transport networks, which are fully utilized across the out-of-sample scenarios. However, small \ce{H2} and \ce{CO2} transport capacities prohibit a quick expansion of the transport infrastructure. Instead, alternative \ce{H2} production technologies have to be deployed, resulting in substantially higher supply chain costs or a failure to achieve the \ce{CO2} emissions targets (\cref{fig:annual_emissions}). 

In contrast, while building large, pan-European \ce{H2} and \ce{CO2} transport networks offers more flexibility, capacities often remain unused if lower \ce{H2} demands materialize, and utilization rates plummet.  

\section{Annual carbon emissions}
\label{supp:co2_emissions}

\begin{figure}[!htp]
	\centering
	{\noindent\includegraphics[width=\textwidth]{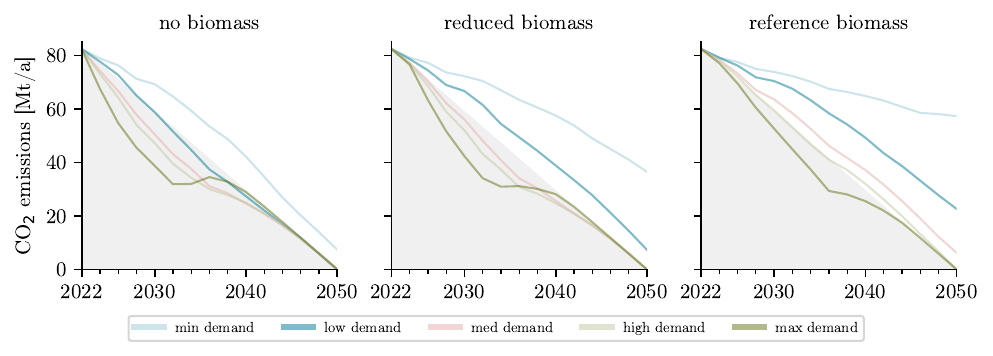}}
	\setlength{\abovecaptionskip}{-15pt}
	\caption{Mean annual \ce{CO2} emissions for each design scenario design.}
	\label{fig:annual_emissions}
\end{figure}

\Cref{fig:annual_emissions} shows the mean annual \ce{CO2} emissions in each design scenario. The grey area visualizes the area in which the annual \ce{CO2} emissions are lower or equal to the annual \ce{CO2} emissions target. In particular, supply chain designs for min and low \ce{H2} demands often exceed the annual \ce{CO2} emission limits, and low-carbon \ce{H2} production has to be substituted with carbonaceous \ce{H2} production to satisfy the \ce{H2} demand. In contrast, supply chains designed to accommodate larger \ce{H2} demands are typically able to adapt their infrastructure quickly enough to achieve the climate targets. Here, we observe that, on average, the HSC emissions stay below the imposed annual \ce{CO2} emission targets. 

\section{Optimistic electrolysis scenario}
\label{suppsec:optimistic_electrolysis}

\begin{figure}[!htp]
	\centering
	{\noindent\includegraphics[width=\textwidth]{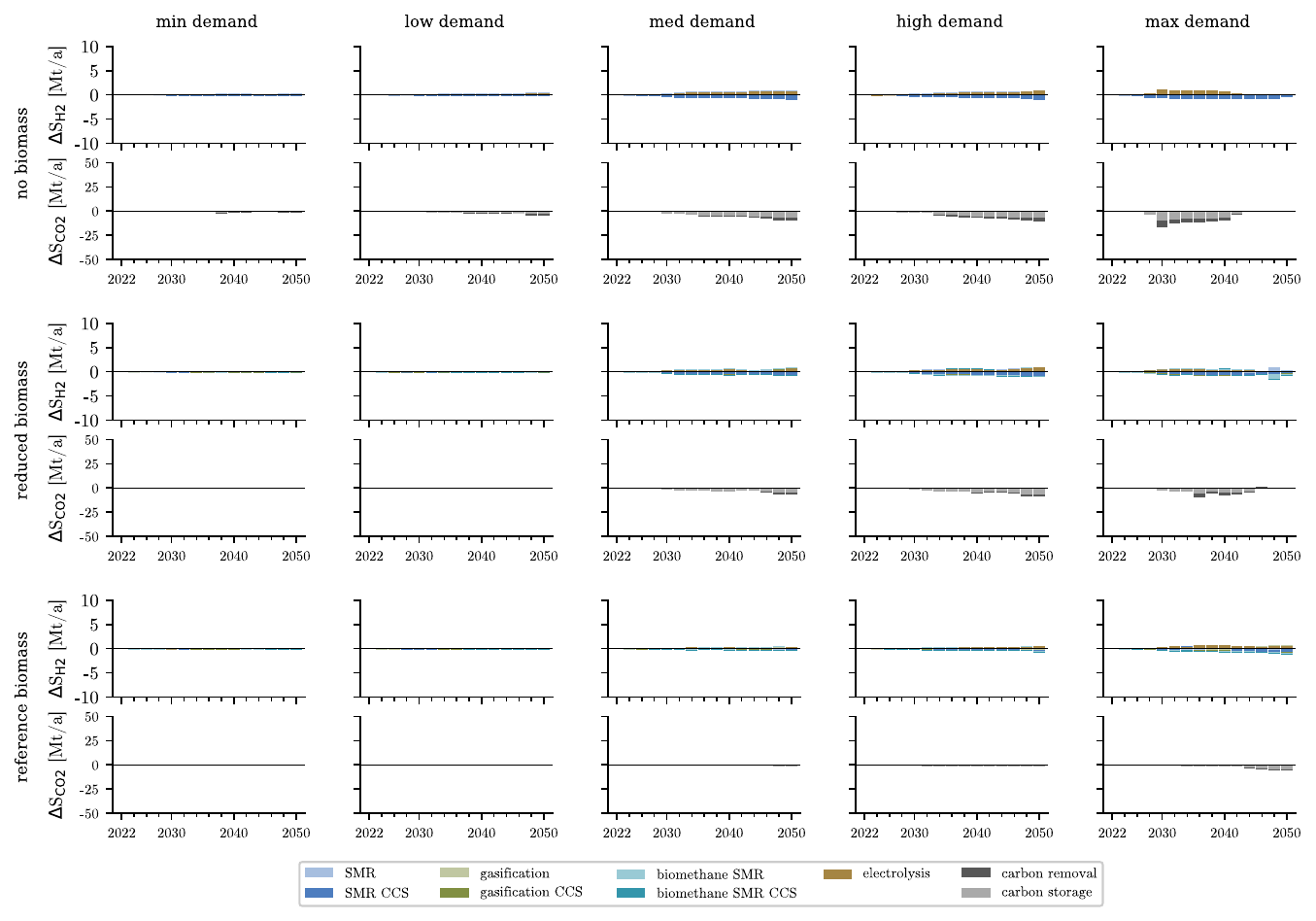}}
	\setlength{\abovecaptionskip}{-15pt}
	\caption{Changes in the cost-optimal \ce{H2} production and \ce{CO2} capture and storage capacities in Mt/a from 2022 to 2050 across all design scenarios for optimistic techno-economic electrolysis assumptions compared to the reference case presented in Fig. 4. \ce{H2} production technologies include steam methane reforming (SMR) from natural gas, biomethane reforming, biomass gasification, and water-electrolysis from electricity. SMR, biomethane reforming, and biomass gasification can be coupled with CCS. In addition, \ce{CO2} removal and \ce{CO2} storage technologies can be installed.}
	\label{fig:evolution_capacities_diff_low_el}
\end{figure}

We include an optimistic case for electrolyzers to investigate the impact of our input parameter assumptions on the investment decisions. To this end, the capital investment cost of the electrolyzers is reduced from 1079\,\euro/kW\textsubscript{\ce{H2}} in 2022 and 413\,\euro/kW\textsubscript{\ce{H2}} in 2050 to 985\,\euro/kW\textsubscript{\ce{H2}} in 2022 and 298\,\euro/kW\textsubscript{\ce{H2}} in 2050. In addition, the electrolysis lifetime is increased from 10 to 20\,years. \Cref{fig:evolution_capacities_diff_low_el} visualizes the changes in the cost-optimal \ce{H2} production capacities between 2022 and 2050 compared to the reference case presented in Fig. 4. Even in this optimistic case, the share of electrolyzers does not increase significantly with respect to the reference case, and changes remain below 8\,\% for the \ce{H2} production capacities. 

\FloatBarrier
\newpage

\section{Optimistic carbon transport cost scenario}
\label{supp:optimistic_CCTS}

\begin{figure}[!htp]
	\centering
	{\noindent\includegraphics[width=\textwidth]{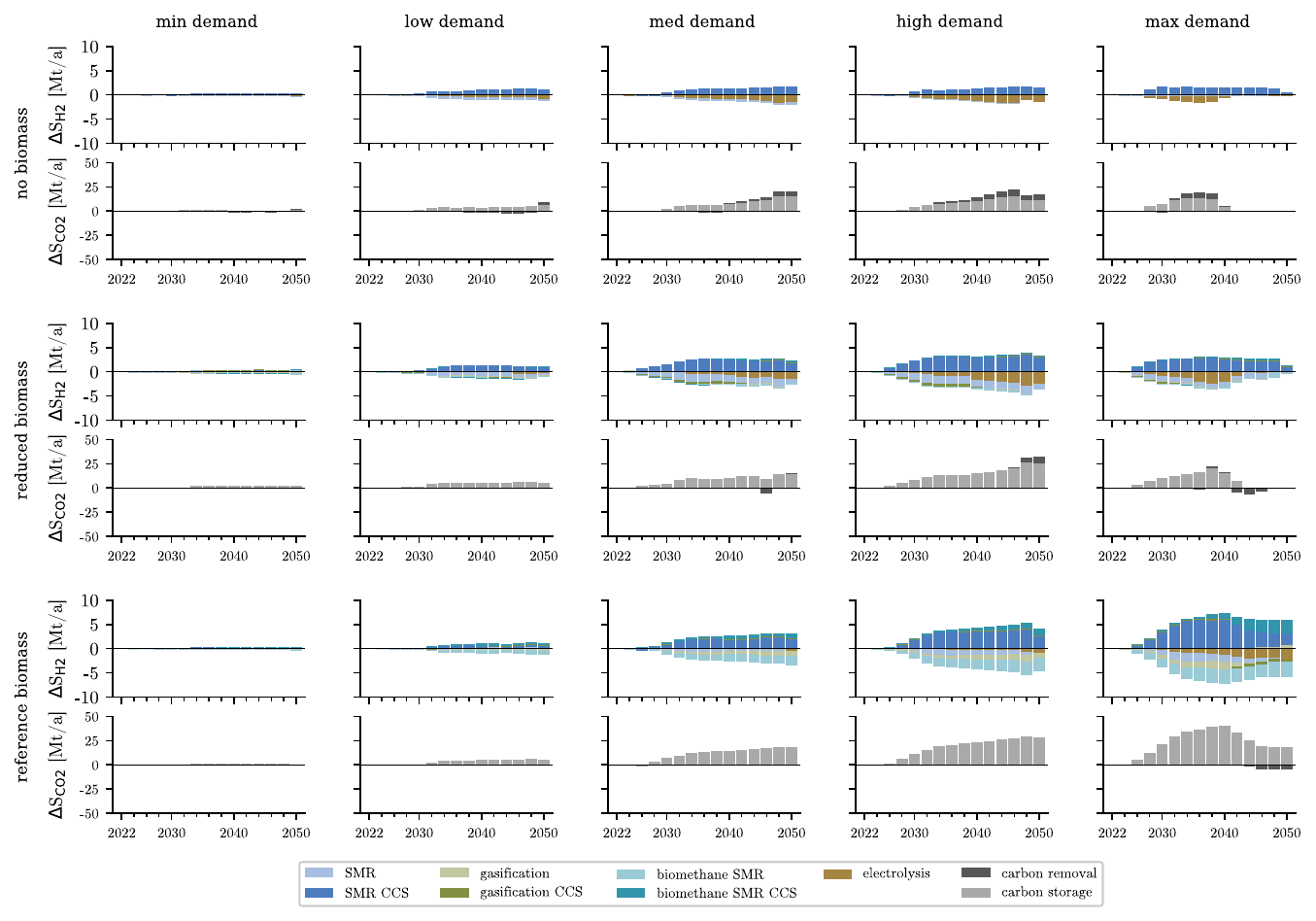}}
	\setlength{\abovecaptionskip}{-15pt}
	\caption{Changes in the cost-optimal \ce{H2} production and \ce{CO2} capture and storage capacities in Mt/a from 2022 to 2050 across all design scenarios for optimistic \ce{CO2} transport cost assumptions compared to the reference case. \ce{H2} production technologies include steam methane reforming (SMR) from natural gas, biomethane reforming, biomass gasification, and water-electrolysis from electricity. SMR, biomethane reforming, and biomass gasification can be coupled with CCS. In addition, \ce{CO2} removal and \ce{CO2} storage technologies can be installed.}
	\label{fig:evolution_capacities_diff_low_CCS}
\end{figure}

We include an optimistic case for \ce{CO2} network to investigate the impact of our input parameter assumptions on the investment decisions. To this end, the capital investment cost of the \ce{CO2} trucks and pipelines is reduced to zero, and only the operational costs for \ce{CO2} trucks (0.5 \euro{}/(t km)) remain. \Cref{fig:evolution_capacities_diff_low_CCS} visualizes the changes in the cost-optimal \ce{H2} production capacities between 2022 and 2050 with respect to the reference case presented in Fig. 4. In design scenarios with biomass availability, we observe a shift from low-carbon \ce{H2} production from biomass-based \ce{H2} production to \ce{H2} production from natural gas coupled with CCS. Design scenario designs with med-max \ce{H2} demands are affected the most, where biomass-based \ce{H2} production capacities reduce by up to 16\,\% while SMR-CCS capacities increase by up to 26\,\%. Furthermore, investments in \ce{CO2} removal and \ce{CO2} storage capacities occur earlier in time, and \ce{CO2} storage capacities increase up to 15\,\% (see reference biomass, max demand). Scenarios that do not include biomass are less affected, and changes remain below 10\,\% and 15\,\% for \ce{H2} production capacities and \ce{CO2} capture and storage capacities, respectively.

\FloatBarrier
\newpage

\section{90\protect\,\protect\% decarbonization scenario}

\begin{figure}[!htp]
	\centering
	{\noindent\includegraphics[width=\textwidth]{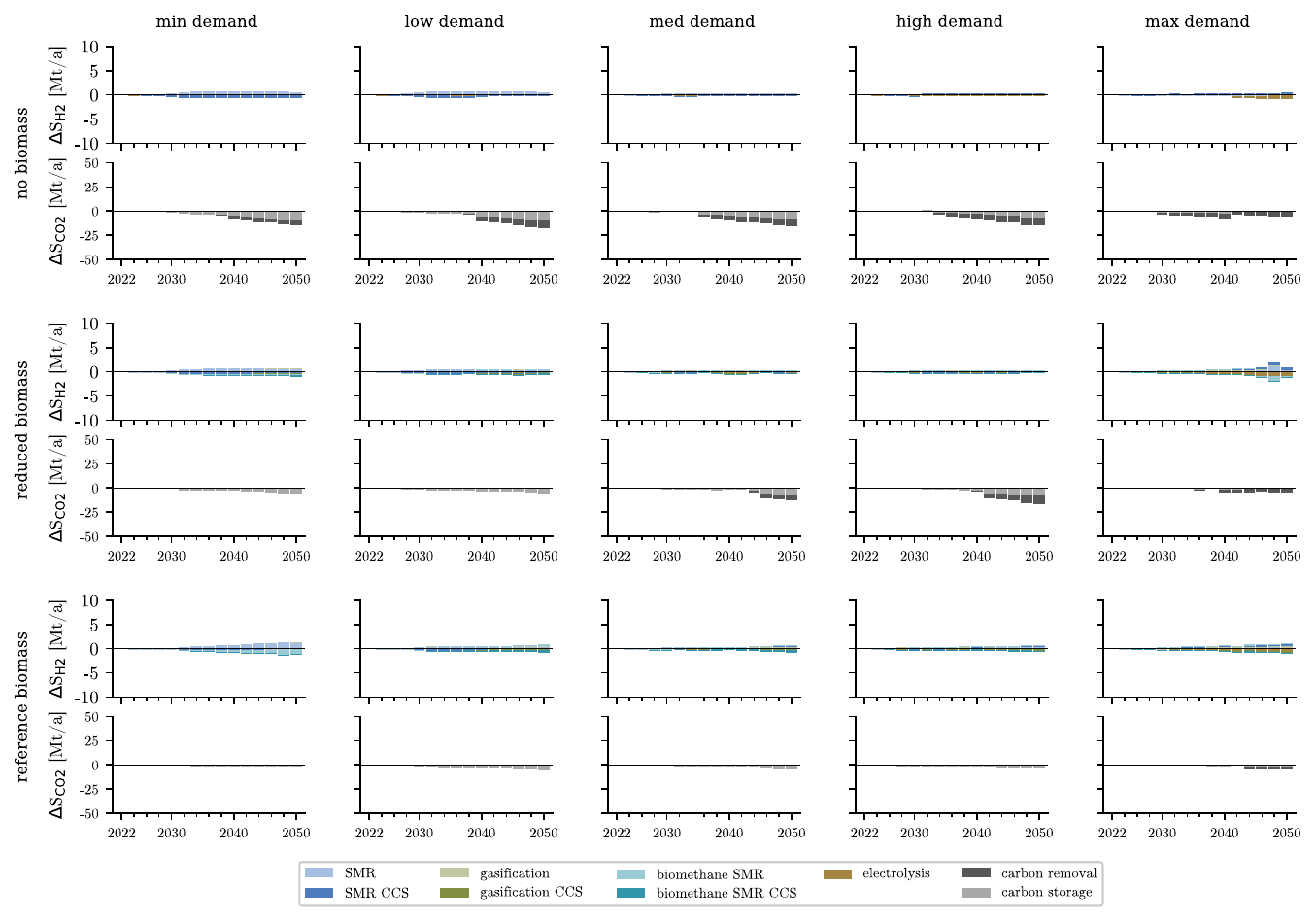}}
	\setlength{\abovecaptionskip}{-15pt}
	\caption{Changes in the cost-optimal \ce{H2} production and \ce{CO2} capture and storage capacities in Mt/a from 2022 to 2050 across all design scenarios for a 90\,\% decarbonization target by 2050 compared to the reference case reported in Fig. 4. \ce{H2} production technologies include steam methane reforming (SMR) from natural gas, biomethane reforming, biomass gasification, and water-electrolysis from electricity. SMR, biomethane reforming, and biomass gasification can be coupled with CCS. In addition, \ce{CO2} removal and \ce{CO2} storage technologies can be installed.}
	\label{fig:evolution_capacities_diff_90}
\end{figure}

We include a case with a reduced decarbonization target of 90\,\% by 2050 to investigate the impact of the net-zero emissions target on our investment decisions. \Cref{fig:evolution_capacities_diff_90} visualizes the changes in the cost-optimal \ce{H2} production capacities between 2022 and 2050 compared to the reference case presented in Fig. 4. We observe that even for a reduced decarbonization target of 90\,\% by 2050, \ce{CO2} capture and storage infrastructure is required. However, on average, \ce{CO2} capture and storage capacities reduce by about 24\,\%.

\clearpage

\bibliographystyle{elsarticle-num} 
\bibliography{references}

\end{document}